\documentclass[twocolumn,trackchanges]{aastex631}
\usepackage{float}

\newcommand{\teff}{$T_{\rm eff}$~}
\newcommand{\teffns}{$T_{\rm eff}$}

\newcommand{\logg}{$\log g$~}
\newcommand{\loggns}{$\log g$}

\usepackage{array}
\received{August 17, 2024}
\revised{November 8, 2024}
\accepted{November 10, 2024}

\shorttitle{Chemical Fingerprints of M Dwarfs}
\shortauthors{Jahandar et al.}
\begin{document}

\title{Chemical Fingerprints of M Dwarfs: High-Resolution Spectroscopy on 31 M Dwarfs with SPIRou}

\correspondingauthor{Farbod Jahandar}
\email{farbod.jahandar@umontreal.ca}

\author[0000-0003-0029-2835]{Farbod Jahandar} 
\affiliation{Trottier Institute for Research on Exoplanets, Département de Physique, Université de Montréal, 1375 Ave Thérèse-Lavoie-Roux, Montréal, QC, H2V 0B3, Canada}

\author[0000-0001-5485-4675]{Ren\'e Doyon} 
\affiliation{Trottier Institute for Research on Exoplanets, Département de Physique, Université de Montréal, 1375 Ave Thérèse-Lavoie-Roux, Montréal, QC, H2V 0B3, Canada}
\affiliation{Observatoire du Mont-M\'egantic, Universit\'e de Montr\'eal, Montr\'eal H3C 3J7, Canada}

\author[0000-0003-3506-5667]{\'Etienne Artigau} 
\affiliation{Trottier Institute for Research on Exoplanets, Département de Physique, Université de Montréal, 1375 Ave Thérèse-Lavoie-Roux, Montréal, QC, H2V 0B3, Canada}
\affiliation{Observatoire du Mont-M\'egantic, Universit\'e de Montr\'eal, Montr\'eal H3C 3J7, Canada}

\author[0000-0003-4166-4121]{Neil J. Cook} 
\affiliation{Trottier Institute for Research on Exoplanets, Département de Physique, Université de Montréal, 1375 Ave Thérèse-Lavoie-Roux, Montréal, QC, H2V 0B3, Canada}

\author[0000-0001-9291-5555]{Charles Cadieux}
\affiliation{Trottier Institute for Research on Exoplanets, Département de Physique, Université de Montréal, 1375 Ave Thérèse-Lavoie-Roux, Montréal, QC, H2V 0B3, Canada}

\author[0000-0001-5541-2887]{Jean-Fran\c cois Donati} 
\affiliation{Universit\'e de Toulouse, CNRS, IRAP, 14 Avenue Belin, 31400 Toulouse, France}

\author[0000-0001-6129-5699]{Nicolas B. Cowan} % CONFIRMED AS CO-AUTHOR
\affiliation{Department of Earth \& Planetary Sciences, McGill University, 3450 rue University, Montréal, QC H3A 0E8, Canada}
\affiliation{Department of Physics, McGill University, 3600 rue University, Montréal, QC H3A 2T8, Canada}

\author[0000-0001-5383-9393]{Ryan Cloutier}
\affiliation{Department of Physics \& Astronomy, McMaster University, 1280 Main St W, Hamilton, ON L8S 4L8, Canada}

\author[0000-0002-8573-805X]{Stefan Pelletier}
\affiliation{Trottier Institute for Research on Exoplanets, Département de Physique, Université de Montréal, 1375 Ave Thérèse-Lavoie-Roux, Montréal, QC, H2V 0B3, Canada}
\affiliation{Observatoire astronomique de l’Université de Genève, 51 chemin Pegasi 1290 Versoix, Switzerland}

\author{Alan Alves-Brito}
\affiliation{Universidade Federal do Rio Grande do Sul, Instituto de Física, Av. Bento Gonçalves 9500, Porto Alegre, RS, Brazil}

\author[0000-0002-1532-9082]{Jorge H. C. Martins} % CONFIRMED AS CO-AUTHOR
\affiliation{Instituto de Astrofísica e Ciências do Espaço, Universidade do Porto, CAUP, Rua das Estrelas, 4150-762, Porto, Portugal}

\author[0000-0001-8385-9838]{Hsien Shang}
\affiliation{Institute of Astronomy and Astrophysics, Academia Sinica,  Taipei 10617, Taiwan}

\author[0000-0003-2471-1299]{Andr\'es Carmona} 
\affiliation{Univ.\ Grenoble Alpes, CNRS, IPAG, 38000 Grenoble, France}

% \author{the SPIRou SLS team}



\begin{abstract}

We extend the methodology introduced by \cite{Jahandar2023} to determine the effective temperature and chemical abundances of 31 slowly-rotating solar neighborhood M dwarfs (M1-M5) using high-resolution spectra from CFHT/SPIRou. This group includes 10\,M dwarfs in binary systems with FGK primaries of known metallicity from optical measurements. By testing our \teff method on various synthetic models, we find a consistent inherent synthetic uncertainty of $\sim$10\,K at a signal-to-noise ratio greater than 100. Additionally, we find that our results align with interferometric measurements, showing a consistent residual of $-$29\,$\pm$\,31\,K. Taking the inherent uncertainties into account, we infer the \teff values of our targets and find an excellent agreement with previous optical and NIR studies. Our high-resolution chemical analysis examines hundreds of absorption lines using $\chi^2$ minimization using PHOENIX-ACES stellar atmosphere models. We present elemental abundances for up to 10 different elements, including refractory elements such as Si, Mg, and Fe, which are important for modelling the interior structure of exoplanets.
In binary systems, we find an average [Fe/H] of $-$0.15\,$\pm$\,0.08 for M dwarfs, marginally lower than the reported metallicity of $-$0.06\,$\pm$\,0.18 for the FGK primaries from \cite{mann2013prospecting}.
We also observe slightly sub-solar chemistry for various elements in our non-binary M dwarfs, most notably for O, C, and K abundances. In particular, we find an average metallicity of $-$0.11\,$\pm$\,0.16 lower but still consistent with the typical solar metallicity of FGK stars (e.g. [Fe/H]\,=\,0.04\,$\pm$\,0.20 from \citealt{brewer2016spectral}). This study highlights significant discrepancies in various major M dwarf surveys likely related to differences in the methodologies employed.

\end{abstract}



\section{Introduction} \label{sec:intro}

M dwarfs, the most populous stellar type in our galaxy, play a pivotal role in astrophysical studies (\citealt{henry1994solar}; \citealt{winters2019solar}; \citealt{Reyle_2021}). Their prevalence and unique properties have made them subjects of intense research in various fields, from galaxy evolution to exoplanet interior modeling. Moreover, M dwarfs are known to host exoplanets (e.g., \citealt{bonfils2013harps}; \citealt{dressing2015occurrence}; \citealt{mulders2015increase}; \citealt{gaidos2016they}; \citealt{Cloutier-Menou_2020}; \citealt{Hsu_2020}), making them vital in the study of different star systems and the search for potentially habitable planets. Accurate measurement of the chemical composition of M dwarfs is necessary to assess the habitability of planets orbiting them.

Various studies have found a direct correlation between the metallicity of host stars and their exoplanets (\citealt{santos2004spectroscopic}; \citealt{fischer2005planet}; \citealt{bond2006abundance}; \citealt{guillot2006correlation}; \citealt{mulders2015increase}). The importance of metallicity extends to the efficiency of giant planet formation, particularly for FGK dwarfs. FGK stars typically possess higher disk masses than M dwarfs, enabling them to form giant planets even with low metallicities. In contrast, M dwarfs require high metallicity to compensate for their lower disk masses to form giant planets (\citealt{neves2013metallicity}).

Furthermore, the chemical composition of a host star has implications for the composition of its planets (\citealt{gilli2006abundances}; \citealt{adibekyan2021composition}). For instance, the relative abundances of refractory elements like Mg, Si, and Fe influence the chemical composition of the protoplanetary disk, more specifically the core-to-mantle mass ratio of rocky exoplanets (\citealt{brewer2016c}). 

A common method for characterizing M dwarf properties is high-resolution spectroscopy (\citealt{souto2017chemical}; \citealt{rajpurohit2018exploring}; \citealt{passegger2019carmenes}; \citealt{marfil2021carmenes}; \citealt{cristofari2022estimating2}; \citealt{Jahandar2023}). However, the spectroscopic analysis of M dwarfs presents several challenges. In contrast to more massive stars like FGK dwarfs, whose spectra are typically dominated by atomic lines, the cooler temperatures of M dwarfs result in dense molecular bands. These bands, such as TiO, VO, and CaH, can overlap and obscure the atomic lines crucial for chemical analysis in the optical regime. In the near-infrared (NIR) regime, additional challenges arise with extensive bands from water vapor, OH bands and hydrides like FeH complicating the identification of atomic lines. The presence of multi-metal molecules like CO and CN adds to this complexity, especially when disentangling each atom's individual contributions to the overall spectral signature. Nevertheless, as the flux of M dwarfs is predominantly in the NIR band, this wavelength regime has become a common focus for chemical spectroscopy on M dwarfs.

M dwarfs in binary systems, particularly when paired with FGK primary stars, present a unique opportunity to better understand their chemical evolution. These binary systems are particularly attractive because they are believed to have formed together from the same molecular cloud, thus sharing similar initial chemical abundances. Due to their higher effective temperature (\teffns), FGK stars are significantly easier to characterize than M dwarfs chemically. Therefore, if they share similar chemistry with their M dwarf companions, it provides a unique opportunity to calibrate spectroscopic methods, which are often subjected to various uncertainties and systematic effects (\citealt{blanco2019modern}; \citealt{olander2021comparative}; \citealt{Jahandar2023}).

In our previous work \citep{Jahandar2023}, we tackled the challenge of minimizing the impact of several caveats observed in synthetic NIR high-resolution atmosphere models used for stellar characterization of M dwarfs.  Our mitigating strategies include a line selection and identifying flaws in these models, such as continuum mismatch, line shift and unresolved contamination. Recognizing and addressing these factors is crucial, as neglecting them can result in inaccurate estimates of a star's fundamental characteristics such as its \teff and metallicity.

Building upon the methodologies and insights established in \cite{Jahandar2023}, this paper presents the chemical spectroscopy of 31 M dwarfs observed by the SPIRou instrument. This includes a high-resolution spectral analysis of hundreds of absorption lines and molecular bands in their spectra and the determination of the \teff and chemical abundances of 10 different elements. 

The paper is structured as follows. We describe the observations with SPIRou in Section \ref{sec:observation}. In Section \ref{sec:spectral_analysis}, we outline our approach for determining \teff and the chemical abundances of our M dwarfs, presenting a fine-tuned line list of those spectral features commonly observed in different stars. In Section \ref{sec:discussion}, we report the chemical abundances of up to 10 different elements for all our M dwarfs and compare them with previously reported values. Moreover, we discuss discrepancies in the molar ratio of refractory elements between this work and others. This analysis is followed by concluding remarks in Section \ref{sec:conclusion}.

\section{Observations} \label{sec:observation}

Our observed spectra were acquired using the Spectro-Polarimètre Infra-Rouge (SPIRou, \citealt{donati2020spirou}). This instrument delivers high-resolution NIR spectra with R\,$\sim$\,70\,000 in the $YJHK$ bands. Here we present high-resolution chemical spectroscopy on two distinct spectra sets from the SPIRou Legacy Survey (SLS) and 10 M dwarfs paired with FGK primaries (PI: Jahandar), the specifics of which will be defined later in the following sections. The telluric corrections and normalization were explained in detail in \cite{Jahandar2023}, but here we briefly explain the procedure. The data were initially reduced using APERO (A PipelinE to Reduce Observations; \citealt{cook2022apero}), which performs calibration, extraction, and telluric corrections on the raw observed data. These corrections yield relative radial velocity (RV) precision at the m/s scale, with residual telluric effects of $<$1\% of the continuum level in the combined spectrutm. Following the telluric correction routine, all spectra were normalized using the iSpec tools (\citealt{blanco2014determining}; \citealt{blanco2019modern}). This tool effectively detects strong spectral features and normalizes the continuum via a cubic spline fitting routine on the whole spectra.

\subsection{Slow Rotating SLS M dwarfs}

The SPIRou Legacy Survey is a large program mostly focused on the detection and characterization of exoplanets, along with magnetic field studies of young stars (\citealt{donati2020spirou}). The primary goal of the SLS survey is to search for potentially habitable exoplanets around nearby M dwarfs using the RV method (\citealt{campbell1988search}; \citealt{latham1989unseen}; \citealt{mayor1995jupiter}). The survey targets a sample of tens of nearby M dwarfs likely to host habitable planets. The high-resolution spectra obtained from the SLS survey can provide valuable information about the properties of M dwarfs. Here, we use 21 high-quality (all with SNR\,$>$\,500) slowly-rotating M dwarfs (\citealt{donati2023magnetic}, hereafter SLS sample; see the upper section of Table \ref{table:mdwarfs_map}). The main reason we chose slow-rotating M dwarfs was to mitigate the potential effects of their high magnetic fields that could cause Zeeman splitting and broadening of spectral lines, which is not included in the synthetic models.

\subsection{M dwarf Binary Systems with FGK Primary}

The second dataset comprises high-resolution spectra of 10 M dwarfs that are in binary systems with an FGK star whose metallicity is well-determined from optical high-resolution spectroscopy. These targets were observed over a two-year program spanning from 2018 to 2020 (see the lower section of Table \ref{table:mdwarfs_map}). Previous work by \cite{mann2013prospecting} on the FGK primaries of these targets focused on developing Equivalent Width (EW)-metallicity calibrations for late-K and M dwarfs. This involved analyzing moderate-resolution optical and NIR spectra from the M dwarfs in 112 binary systems, each paired with a high-resolution optically studied FGK primary star. Their methodology centered on correlating the EW of metal-sensitive features in M dwarfs' spectra, successfully establishing a relationship between EW measurements in M dwarfs and the metallicities of their FGK primaries.

The main reason for including these M dwarfs in our dataset is to determine their metallicity directly using only atomic lines in their high-resolution NIR spectra, independent of their FGK primaries. This approach allows us to refine and calibrate chemical spectroscopic methods for M dwarfs.

\begin{deluxetable}{lccccl}[ht]
\tablecaption{Selected slow rotating M dwarfs}
\tabletypesize{\scriptsize} 
\tablehead{
\colhead{Name} & \colhead{Visits} & \colhead{M-SNR$^{\dag}$} & \colhead{CA-SNR$^{*}$} & \colhead{Obj} & \colhead{Primary}}

\startdata
V* GX And &  1283 &      271 &    1107.4  & SLS & ~~~~~---\\
L 675-81  &   104 &       98 &    565.4  & SLS & ~~~~~---\\
HD 95735  &   521 &      324 &    1622.1  & SLS & ~~~~~---\\
Ross 104  &   523 &      123 &    1093.0  & SLS & ~~~~~---\\
BD+68 946 &   894 &      180 &    1233.6  & SLS & ~~~~~---\\
HD 265866 &   630 &      130 &    1084.1  & SLS & ~~~~~---\\
G 31-53   &   449 &       91 &    865.7  & SLS & ~~~~~--- \\
G 68-8    &   703 &       93 &    965.8  & SLS & ~~~~~---\\
BD-15 6290 &   371 &      154 &    823.0  & SLS & ~~~~~---\\
Ross 1003 &   315 &       88 &    768.6  & SLS & ~~~~~--- \\
HD 173740 &   856 &      151 &    1043.9  & SLS & ~~~~~---\\
G 111-47  &   440 &       94 &    961.9  & SLS & ~~~~~--- \\
G 192-13  &   603 &       96 &    923.7  & SLS & ~~~~~---\\
G 254-29  &   196 &       96 &    822.8  & SLS & ~~~~~---\\
 GJ 699   &   893 &      182 &    1458.7  & SLS & ~~~~~---\\
V* GQ And &   731 &       95 &    961.2  & SLS & ~~~~~---\\
G 130-4   &   743 &       85 &    1020.0  & SLS & ~~~~~---\\
Ross 248 &   718 &      104 &    1078.0  & SLS & ~~~~~---\\
G 122-49  &   391 &       84 &    739.8  & SLS & ~~~~~--- \\
G 157-77  &   413 &       76 &    639.5  & SLS & ~~~~~---\\
G 158-27  &   451 &       90 &    794.3  & SLS & ~~~~~--- \\
 \hline
 \hline
HD 154363B  &        4 &      111 &          189.0 &   Bin & HIP 83591\\
G 232-62    &        4 &       51 &          97.7 &   Bin & HIP 109926\\
HD 263175B  &        4 &       54 &          105.3 &   Bin & HIP 32423\\
* 20 LMi B  &        4 &       42 &           75.4 &   Bin & HIP 49081\\
HD 183870B  &        4 &       51 &          98.6 &   Bin & HIP 96085\\
LP 128-32   &        4 &       31 &           54.9 &   Bin & HIP 53008\\
NLTT 39578  &        4 &       39 &           78.1 &   Bin & HIP 74396\\
NLTT 44569  &        4 &       47 &           89.4 &   Bin & HIP 84616\\
G 275-2     &        4 &       51 &          99.7 &   Bin & HIP 114156\\
G 106-36    &       12 &       37 &          92.3 &   Bin & HIP 29860\\
\enddata
\tablecomments{
$^{\dag}$ Mean SNR at the $H$ band\\
$^{*}$ Co-Added SNR at the $H$ band\\
}
\label{table:mdwarfs_map}
\end{deluxetable}

\section{Spectral Analysis} \label{sec:spectral_analysis}

For the spectral fitting, we use pre-generated synthetic spectra from the PHOENIX-ACES grid (\citealt{allard2012models}; \citealt{husser2013new}). Initially, we determine the \teff by performing bulk line-by-line fitting on hundreds of lines per spectrum, using the methodology of \cite{Jahandar2023} to estimate the \teff and overall metallicity ([M/H]) of our M dwarfs. The methodology is briefly summarized here. To minimize discrepancies often seen between models and observed data, like the continuum mismatch, we select only those spectral features that closely align with our models. These lines are then grouped into sets of several lines each and individually analyzed using a $\chi^2$ minimization method, providing multiple independent measurements (see Figures 8 and 9 in \citealt{Jahandar2023}), whose dispersion provides an uncertainty of \teff for each target.  This method effectively yields the best overall metallicity together with the optimal \teffns.

The \logg values of our targets range between 4.8 and 5.2 dex, as estimated from the TESS database \citep{stassun2019revised}. We found that a variation of 5.0\,$\pm$\,0.2 dex does not significantly impact our results, leading us to use a constant \logg of 5.0 dex for all targets. Initially, we obtained the pre-generated synthetic spectra with \teff values ranging from 2300\,K to 4300\,K and metallicity between $-1.0$ and 1.0 dex, with 100\,K and 0.5 dex increments. However, these incremental spaces were too coarse for our purpose. Therefore, we performed bilinear interpolation on our synthetic grid to achieve finer incremental ranges of 20\,K and 0.1 dex. The PHOENIX-ACES models for M dwarfs constrain us to adjust only the overall metallicity, \loggns, and \teff with a preset [$\alpha$/M]. Given that our study does not quantify individual elemental abundances in molecules such as CO and CN, which necessitates simultaneous control of both [M/H] and [$\alpha$/M], the fixed [$\alpha$/M] assumption does not significantly affect our analysis of atomic lines.

For determining the chemical abundances, using our estimated \teffns, we adjust the synthetic spectra's metallicity to find the best match for each atomic or molecular line (see Table \ref{table:abundances}). By fixing \teff and \logg values, we mitigate the risk of extra uncertainty that could arise from similar line shapes produced by different combinations of \teffns, \loggns, and metallicity.

\begin{deluxetable*}{lcrrrrrrrrrrr}
\rotate
\tabletypesize{\scriptsize} 
\tablecaption{Solar normalized elemental abundances of our M dwarf targets}
\tablehead{
     \colhead{Names} &  \colhead{Teff} & \colhead{[M/H]$^*$} & \colhead{[Fe/H]} & \colhead{[Mg/H]} & \colhead{[O/H]$~\hat{}\ $} & \colhead{[Si/H]}  & \colhead{[Ca/H]} & \colhead{[Ti/H]}  & \colhead{[Al/H]} & \colhead{[Na/H]}  & \colhead{[C/H]}  & \colhead{[K/H]}} 
\startdata
          V* GX And &  3661$\pm$27 &  $-$0.35$\pm$0.03 &   $-$0.39$\pm$0.03 &   $-$0.24$\pm$0.03 & $-$0.48$\pm$0.02  &  $-$0.36$\pm$0.03 &  $-$0.50$\pm$0.08   &  $-$0.19$\pm$0.03 &  $-$0.26$\pm$0.03 &  $-$0.30$\pm$0.08 &  $-$0.35$\pm$0.05    &  -0.40 $\pm$ 0.08  \\
          L  675-81 &  3497$\pm$54 &   0.19$\pm$0.08   &    0.07$\pm$0.04   &    0.05$\pm$0.04   & $-$0.23$\pm$0.03  &    0.45$\pm$0.05  &  $-$0.04$\pm$0.03   &    0.19$\pm$0.04  &    0.10$\pm$0.07  &    0.55$\pm$0.05  &   0.35$\pm$0.05      &   0.43 $\pm$ 0.04  \\
          HD  95735 &  3593$\pm$23 &  $-$0.33$\pm$0.05 &   $-$0.27$\pm$0.05 &   $-$0.13$\pm$0.03 & $-$0.23$\pm$0.02  &  $-$0.35$\pm$0.05 &  $-$0.63$\pm$0.03   &  $-$0.22$\pm$0.02 &  $-$0.14$\pm$0.06 &  $-$0.50$\pm$0.08 &  $-$0.31$\pm$0.15    &  -0.53 $\pm$ 0.03  \\
          Ross  104 &  3483$\pm$35 &  $-$0.09$\pm$0.05 &   $-$0.01$\pm$0.04 &    0.03$\pm$0.04   & $-$0.31$\pm$0.03  &    0.03$\pm$0.05  &  $-$0.30$\pm$0.12   &    0.03$\pm$0.03  &    0.00$\pm$0.04  &  $-$0.15$\pm$0.05 &   0.00$\pm$0.04      &  -0.20 $\pm$ 0.12  \\
        BD+68   946 &  3458$\pm$36 &   0.04$\pm$0.06   &    0.08$\pm$0.05   &    0.31$\pm$0.03   & $-$0.07$\pm$0.03  &    0.15$\pm$0.10  &  $-$0.30$\pm$0.12   &    0.14$\pm$0.05  &    0.15$\pm$0.05  &  $-$0.07$\pm$0.03 &   0.15$\pm$0.05      &  -0.16 $\pm$ 0.03  \\
          HD 265866 &  3422$\pm$46 &   0.02$\pm$0.05   &    0.04$\pm$0.05   &    0.16$\pm$0.03   & $-$0.31$\pm$0.03  &    0.20$\pm$0.06  &  $-$0.20$\pm$0.12   &    0.07$\pm$0.06  &    0.13$\pm$0.05  &    0.10$\pm$0.06  &   0.01$\pm$0.08      &   0.00 $\pm$ 0.12  \\
           G  31-53 &  3385$\pm$47 &   0.06$\pm$0.05   &    0.04$\pm$0.04   &    0.28$\pm$0.04   & $-$0.10$\pm$0.03  &    0.02$\pm$0.08  &  $-$0.19$\pm$0.01   &    0.14$\pm$0.03  &    0.20$\pm$0.06  &    0.13$\pm$0.07  &   ---                &   0.00 $\pm$ 0.08  \\
            G  68-8 &  3349$\pm$17 &   0.33$\pm$0.08   &    0.21$\pm$0.05   &    0.43$\pm$0.03   & $-$0.11$\pm$0.02  &    0.21$\pm$0.11  &    0.00$\pm$0.08    &    0.45$\pm$0.06  &    0.47$\pm$0.07  &    0.50$\pm$0.10  &   0.66$\pm$0.08      &   0.40 $\pm$ 0.07  \\
        BD-15  6290 &  3349$\pm$51 &   0.22$\pm$0.07   &    0.15$\pm$0.06   &    0.30$\pm$0.07   & $-$0.25$\pm$0.04  &    0.15$\pm$0.15  &  $-$0.07$\pm$0.03   &    0.30$\pm$0.07  &    0.30$\pm$0.08  &    0.41$\pm$0.01  &   0.44$\pm$0.16      &   0.43 $\pm$ 0.07  \\
          Ross 1003 &  3340$\pm$42 &   0.15$\pm$0.05   &    0.13$\pm$0.05   &    0.42$\pm$0.03   & $-$0.05$\pm$0.04  &    0.09$\pm$0.06  &  $-$0.10$\pm$0.16   &    0.23$\pm$0.05  &    0.30$\pm$0.10  &    0.20$\pm$0.10  &   0.26$\pm$0.18      &   0.00 $\pm$ 0.16  \\
          HD 173740 &  3325$\pm$25 &  $-$0.25$\pm$0.07 &   $-$0.16$\pm$0.03 &    0.00$\pm$0.12   & $-$0.16$\pm$0.03  &  $-$0.20$\pm$0.12  &  $-$0.47$\pm$0.03  &  $-$0.28$\pm$0.06 &  $-$0.20$\pm$0.12 &  $-$0.45$\pm$0.05 &   0.05$\pm$0.25      &  -0.60 $\pm$ 0.12  \\
           G 111-47 &  3314$\pm$37 &   0.03$\pm$0.07   &   $-$0.05$\pm$0.06 &    0.10$\pm$0.09   & $-$0.29$\pm$0.05  &   ---                 &  $-$0.17$\pm$0.03   &    0.03$\pm$0.07  &    0.15$\pm$0.06  &    0.30$\pm$0.06  &   ---            &   0.19 $\pm$ 0.10  \\
           G 192-13 &  3283$\pm$39 &   $-$0.03$\pm$0.04   &    0.02$\pm$0.04   &    0.04$\pm$0.07   & $-$0.23$\pm$0.04  &    0.10$\pm$0.16   &  $-$0.20$\pm$0.16  &  $-$0.12$\pm$0.06 &    0.03$\pm$0.07  &    0.20$\pm$0.16  &  $-$0.03$\pm$0.04   &  -0.10 $\pm$ 0.16  \\
           G 254-29 &  3325$\pm$21 &  $-$0.25$\pm$0.07 &   $-$0.08$\pm$0.04 &   $-$0.10$\pm$0.09 & $-$0.17$\pm$0.02  &  $-$0.50$\pm$0.08  &  $-$0.40$\pm$0.08  &  $-$0.17$\pm$0.08 &  $-$0.05$\pm$0.09 &  $-$0.45$\pm$0.05 &   0.03$\pm$0.23      &  -0.60 $\pm$ 0.08  \\
           GJ 699   &  3232$\pm$21 &  $-$0.51$\pm$0.04 &   $-$0.34$\pm$0.04 &   ---                  & $-$0.34$\pm$0.04  &  $-$0.50$\pm$0.16 &  $-$0.60$\pm$0.16   &  $-$0.53$\pm$0.05 &  $-$0.47$\pm$0.09 &  $-$0.50$\pm$0.16 &  $-$0.55$\pm$0.05 &  -0.75 $\pm$ 0.05  \\
          V* GQ And &  3237$\pm$14 &  $-$0.47$\pm$0.04 &   $-$0.44$\pm$0.05 &   $-$0.70$\pm$0.10 & $-$0.42$\pm$0.05  &  $-$0.46$\pm$0.16  &  $-$0.50$\pm$0.16  &  $-$0.40$\pm$0.09 &  $-$0.40$\pm$0.06 &  $-$0.40$\pm$0.11 &  $-$0.32$\pm$0.14    &  -0.70 $\pm$ 0.16  \\
            G 130-4 &  3198$\pm$26 &   0.01$\pm$0.05 &   $-$0.09$\pm$0.05 &   $-$0.15$\pm$0.17 & $-$0.23$\pm$0.04  &   ---                 &  $-$0.25$\pm$0.05   &  $-$0.26$\pm$0.15 &  $-$0.07$\pm$0.03 &    0.10$\pm$0.10  &   0.04$\pm$0.04  &   0.15 $\pm$ 0.15  \\
          Ross  248 &  3123$\pm$17 &  $-$0.26$\pm$0.03 &   $-$0.22$\pm$0.09 &   $-$0.30$\pm$0.10 & $-$0.23$\pm$0.09  &   ---                 &  $-$0.35$\pm$0.05   &   ---                 &  $-$0.20$\pm$0.06 &   ---                 &   ---    &   ---   \\
           G 122-49 &  3149$\pm$35 &  $-$0.14$\pm$0.05 &   $-$0.13$\pm$0.04 &   $-$0.24$\pm$0.16  & $-$0.23$\pm$0.04  &    0.00$\pm$0.16   &  $-$0.23$\pm$0.03 &  $-$0.43$\pm$0.07 &    0.00$\pm$0.16  &    0.10$\pm$0.10  &  $-$0.05$\pm$0.13    &  -0.22 $\pm$ 0.16  \\
           G 157-77 &  3056$\pm$36 &  $-$0.24$\pm$0.03 &   $-$0.20$\pm$0.06 &   ---                  &  $-$0.17$\pm$0.02 &   ---                 &   ---                   &  $-$0.20$\pm$0.10 &  $-$0.35$\pm$0.05 &  $-$0.30$\pm$0.08 &   ---    &   ---   \\
           G 158-27 &  3003$\pm$45 &  $-$0.17$\pm$0.07 &   $-$0.18$\pm$0.11 &   ---                  & $-$0.11$\pm$0.03  &   ---                 &   ---                   &  $-$0.15$\pm$0.09 &  $-$0.40$\pm$0.12 &   --- &   0.00$\pm$0.12      &   ---   \\
        \\
    \hline
    \hline
    \\
         HD 154363B &  3633$\pm$20  &  $-$0.08$\pm$0.10 &   $-$0.11$\pm$0.06 &    0.33$\pm$0.05   &   0.01$\pm$0.03  &   ---                 &  $-$0.46$\pm$0.03 &    0.09$\pm$0.05   &   0.15$\pm$0.05   &  $-$0.40$\pm$0.12  &  ---                 & -0.35 $\pm$ 0.05\\
           G 232-62 &  3265$\pm$22  &  $-$0.14$\pm$0.04 &   $-$0.10$\pm$0.11 &  $-$0.22$\pm$0.02  & $-$0.09$\pm$0.04 &   ---                 &  $-$0.20$\pm$0.19 &  $-$0.31$\pm$0.16  & $-$0.05$\pm$0.05  &    0.02$\pm$0.16   &  ---                 & --- \\
         HD 263175B &  3642$\pm$31  &  $-$0.30$\pm$0.06 &   $-$0.27$\pm$0.08 &  $-$0.02$\pm$0.05  & $-$0.43$\pm$0.03 &  $-$0.26$\pm$0.12 &  $-$0.40$\pm$0.07 &  $-$0.22$\pm$0.07  & $-$0.22$\pm$0.01  &   ---                  &  ---                 & -0.54 $\pm$ 0.04\\
        *  20 LMi B &  3106$\pm$54  &  $-$0.34$\pm$0.16 &      0.00$\pm$0.09 &   ---                  & $-$0.13$\pm$0.07 &  $-$0.63$\pm$0.20 &   ---                 &   ---                  & $-$0.60$\pm$0.16  &   ---                  &  ---                 & ---\\
         HD 183870B &  3183$\pm$12  &  $-$0.26$\pm$0.06 &   $-$0.24$\pm$0.06 &   ---                  & $-$0.38$\pm$0.07 &  $-$0.15$\pm$0.16 &  $-$0.15$\pm$0.15 &   ---                  & $-$0.30$\pm$0.16  &  $-$0.10$\pm$0.16  &  ---                 & -0.50 $\pm$ 0.15\\
         LP  128-32 &  3280$\pm$21  &   0.02$\pm$0.13   &   $-$0.10$\pm$0.04 &   ---                  & $-$0.26$\pm$0.04 &    0.20\,$\pm$\ ,0.16 &   ---                 &    0.59$\pm$0.16   & $-$0.18$\pm$0.16  &   ---                  & $-$0.10$\pm$0.16 &  --- \\
         LP  222-50 &  3456$\pm$36  &  $-$0.16$\pm$0.07 &   $-$0.19$\pm$0.07 &    0.05$\pm$0.10   & $-$0.31$\pm$0.02 &  $-$0.21$\pm$0.08 &  $-$0.19$\pm$0.21 &    0.21$\pm$0.14   & $-$0.20$\pm$0.08  &   ---                  &  ---                 & -0.43 $\pm$ 0.03\\
         LP  138-36 &  3224$\pm$15  &  $-$0.41$\pm$0.05 &   $-$0.23$\pm$0.09 &  $-$0.22$\pm$0.20   & $-$0.49$\pm$0.05 &   ---                 &  $-$0.27$\pm$0.09 &  $-$0.48$\pm$0.20  & $-$0.59$\pm$0.01  &   ---                  & $-$0.33$\pm$0.20 & -0.70 $\pm$ 0.20 \\
            G 275-2 &  3355$\pm$32  &  $-$0.14$\pm$0.05 &   $-$0.10$\pm$0.07 &  $-$0.18$\pm$0.10  & $-$0.35$\pm$0.04 &   ---                 &  $-$0.18$\pm$0.02 &    0.09$\pm$0.18   & $-$0.20$\pm$0.16  &    0.02$\pm$0.16   & $-$0.20$\pm$0.16 & -0.20 $\pm$ 0.16 \\
           G 106-36 &  3230$\pm$18  &  $-$0.17$\pm$0.07 &   ---                  &   $-$0.28$\pm$0.17 & $-$0.32$\pm$0.03 &    0.27$\pm$0.12  &  $-$0.16$\pm$0.09 &  $-$0.34$\pm$0.12  & $-$0.17$\pm$0.12  &    0.02$\pm$0.12   & $-$0.20$\pm$0.12 & -0.32 $\pm$ 0.03\\
\enddata
\tablecomments{
$^*$Average abundance of all elements.\\
$\hat{}\ $The oxygen abundance is inferred from OH lines.}
\label{table:abundances}
\end{deluxetable*}

\subsection{Effective Temperature Uncertainty and Calibration} \label{sec:teff_calibration}

Our \teff method has inherent uncertainties associated with our synthetic models that can be quantified, in particular, to determine at which SNR the \teff uncertainty does not improve. One can quantify this uncertainty by applying our \teff method to an ensemble of synthetic spectra of various \teffns, metallicity and SNR level (50\,$<$\,SNR\,$<$\,1000).  This exercise shows that the \teff uncertainty saturates to $\sim$10\,K for SNR exceeding $\sim$100. This uncertainty is a minimum noise floor since in practice stars are significantly different than models. One can estimate this noise floor empirically by taking the median of all \teff uncertainties from the SLS sample whose stars have SNR well above 100. The data from Table \ref{table:abundances} suggest a noise floor of $\sim$30\,K.   

\begin{figure}[ht]
    \centering
    \includegraphics[width=1\linewidth]{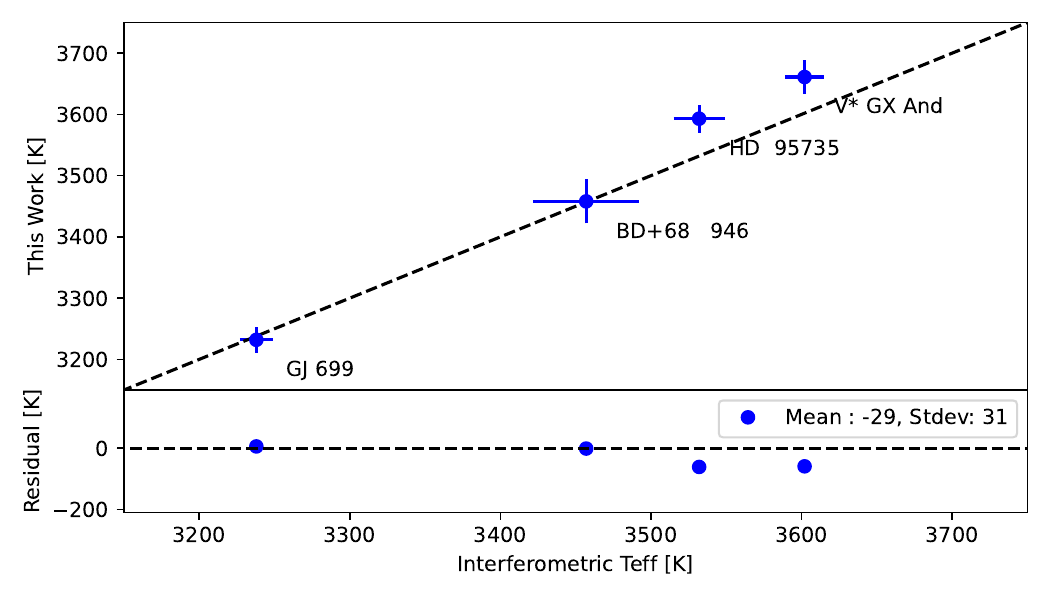}
    \caption{Comparison of the \teff values determined in this study versus interferometric \teff from \cite{mann2013spectro}. The residual plot indicates a mean difference and standard deviation of $-$29\,$\pm$\,31\,K.}
    \label{fig:teff_interf}
\end{figure}

The 30\,K uncertainty mentioned above is internal to our method, but one cannot exclude the possibility that our \teff values are affected by a systematic offset. This hypothesis can be tested by comparing our measurements with stars of known \teff derived from interferometric measurements. The interferometric method is arguably the best model-independent way of determining \teffns, as it only requires the use of the angular diameter and bolometric flux of a target in the Stefan–Boltzmann equation. Our sample comprises four stars with \teff determined through interferometry (from \citealt{mann2013spectro}) that can be used to calibrate our \teff measurements. As shown in Figure \ref{fig:teff_interf}, there is excellent agreement between our method and the interferometric measurements\footnote{Note that HD 173740 also had interferometric measurements in \cite{mann2013spectro}; however, since this work has noted that the photometric or spectroscopic values of this target used in their work are likely inaccurate, which may affect the value they determined, we have removed this target from our comparisons.}. The residual difference between our study and the findings of \cite{mann2013spectro} indicates a marginal offset of $-$29\,$\pm$\,31\,K. While more stars would be needed for better statistics, this comparison confirms the reliability of our method for measuring \teff from NIR high-resolution spectroscopy with absolute uncertainties close to $\sim$30\,K.

\subsection{Line list Fine-tuning}

For this work, we used the most recent available line list from PHOENIX Bt-Settl models (\citealt{allard2010model}) and the atomic database of the National Institute of Standards and Technology, \cite{NISTAtomicSpectraDB}. Throughout our chemical spectroscopy analysis, we observed that some spectral features are common across different M dwarfs, while others are unique to specific stars. To refine our approach, we investigated the frequency of certain lines' visibility across different stars. This assists us in two significant ways: 1) identifying lines that disappear under certain \teff and/or metallicity conditions, and 2) enabling the creation of a fine-tuned line list, which improves the accuracy of future chemical spectroscopy by tailoring line lists to the \teff and metallicity of target stars.

We explicitly avoided including spectral features that appear in only a few stars, as their rare occurrence could indicate a false detection rather than a real spectral line, particularly if not observed in other similar stars. To be more conservative and ensure reliability, we chose only those spectral features that appear in at least five different stars.
The resulting fine-tuned line list is presented in Table \ref{table:line_critera} in the Appendix \ref{sec:appendix2}.

\section{Results and Discussion} \label{sec:discussion}

We have shown that our \teff method can yield reliable measurements with typical absolute uncertainties of $\sim$30\,K. In the following sections, we compare our spectroscopic results (\teff and metallicity) of the binary and the SLS samples to several other spectroscopic works.

\subsection{M dwarfs\,-\,FGK Binary systems}

The continuum of M dwarfs is heavily affected by molecular bands, which complicates line-by-line chemical spectroscopy. A common method for determining the chemical abundances of M dwarfs is to formulate calibration equations based on M dwarfs that have FGK primaries. The fundamental assumption is that both FGK stars and their M dwarf companions originate from the same molecular cloud and, therefore, should share very similar compositions. Studies such as those by \cite{rojas2012metallicity} and \cite{mann2013prospecting} have significantly contributed to this field. These works link the EW of metal-sensitive spectral features, measured from low- and medium-resolution optical and NIR spectra of M dwarfs, to the metallicities derived from their FGK primaries, which are in most cases inferred through high-resolution optical spectroscopy.

\begin{figure*}[!ht]
    \centering
    \includegraphics[width=0.8\linewidth]{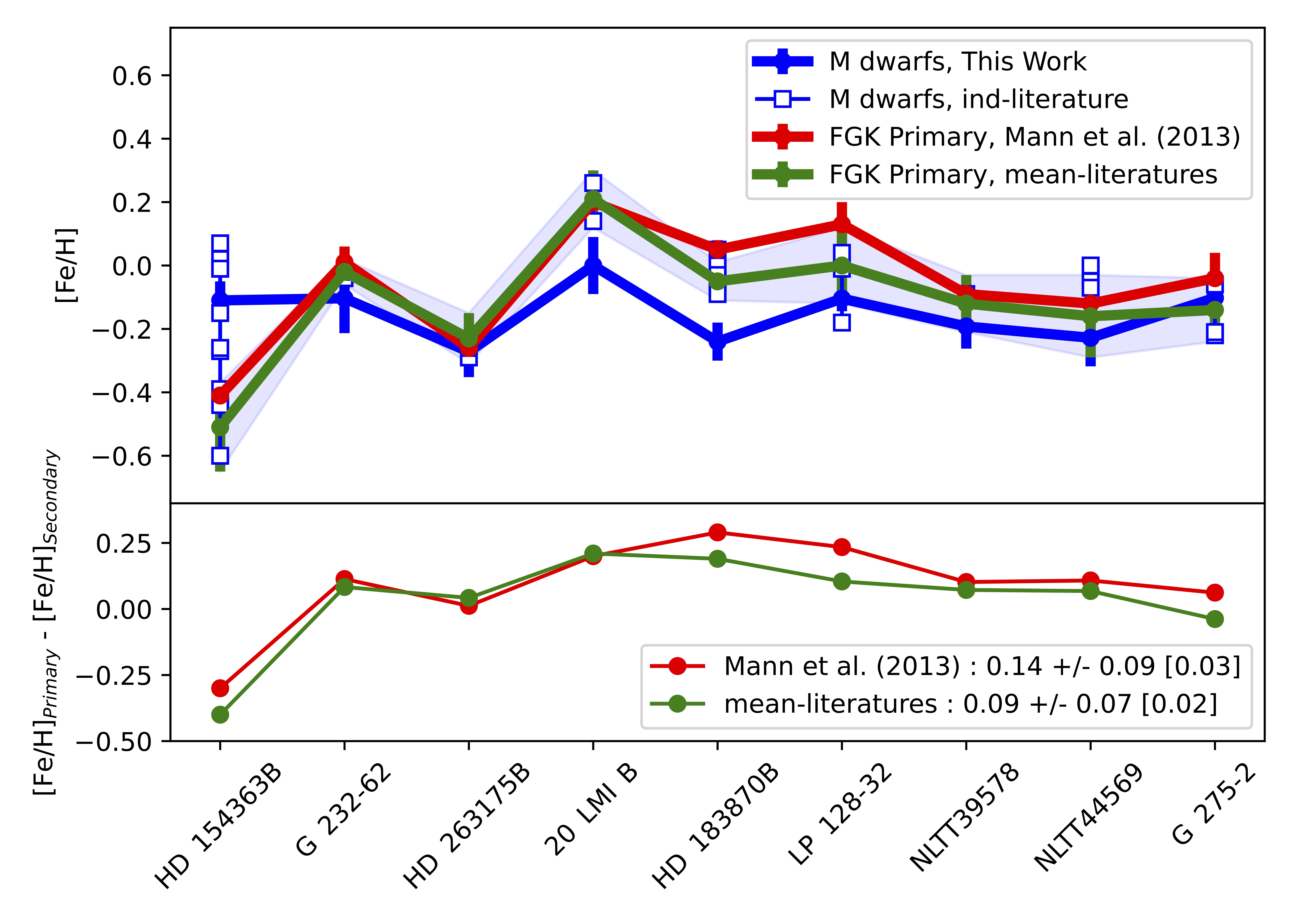}
    \caption{Comparison of the [Fe/H] values of the target M dwarfs from this work with the metallicity of their FGK primaries from other references. \textit{Top panel}: The blue line represents our [Fe/H] for the M dwarf companions, while the red line represents the metallicity for the FGK primaries from \cite{mann2013prospecting}. Each FGK star has at least two metallicity values reported in SIMBAD (\citealt{wenger2000simbad}). The green line represents the averages of the reported metallicities for each star, and the shaded area shows the standard deviation of these values. These values are compiled from the following sources: \citealt{Gonzalez_2008},
    \cite{Da_Silva_2010},
    \cite{Lee_2011},
    \cite{Mishenina_2012},
    \cite{Koleva_2012},
    \cite{Tsantaki_2013},
    \cite{mann2013prospecting},
    \cite{Morel_2013},
    \cite{Mishenina_2013},
    \cite{Gomes_da_Silva_2014},
    \cite{da_Silva_2015},
    \cite{Delgado_Mena_2015},
    \cite{Boeche_2016},
    \cite{brewer2016spectral},
    \cite{Houdebine_2016},
    \cite{Kim_2016},
    \cite{Luck_2016},
    \cite{Delgado_Mena_2017},
    \cite{Aguilera_G_mez_2018},
    \cite{Montes_2018},
    \cite{Soto_2018},
    \cite{Grieves_2018},
    \cite{Arentsen_2019},
    \cite{Morris_2019},
    \cite{Casali_2020},
    \cite{Rice_2020},
    \cite{Liu_2020},
    \cite{J_nsson_2020},
    \cite{Rosenthal_2021} and
    \cite{Hirsch_2021}. The blue empty squares represent other independent measurements of M dwarfs metallicities in the literature, compiled from: \cite{gaidos2014trumpeting}, \cite{mann2015constrain}, \cite{Newton_2018}, \cite{kuznetsov2019characterization}, \cite{schweitzer2019carmenes}, \cite{passegger2019carmenes} and \cite{marfil2021carmenes}. \textit{Bottom panel}: The residual plot shows the differences between the [Fe/H] of M dwarfs from this work and that of the FGK primaries. Our results are slightly more consistent with the averaged literature values.
    }
    \label{fig:metallicity_binaries}
\end{figure*}

With recent developments in high-resolution spectrographs like SPIRou, astronomers can now measure individual abundances from isolated atomic lines in M dwarfs (\citealt{souto2017chemical}; \citealt{ishikawa2022elemental}; \citealt{hejazi2023elemental}). We have determined [Fe/H] from several individual Fe I lines for 10 M dwarfs with FGK primaries. Then we compared the measured [Fe/H] of these M dwarfs with the metallicities of their FGK primaries, as reported by \cite{mann2013prospecting}\footnote{Note that the values reported by \cite{mann2013prospecting} are a collection of the metallicities derived from their work and other studies, including those by \cite{valenti2005spectroscopic}, \cite{robinson2007n2k} and \cite{fuhrmann2008nearby}.}, and with the average of abundances from the literature (see Figure \ref{fig:metallicity_binaries}). Additionally, we over-plotted independent metallicity measurements of the same M dwarfs from different studies to illustrate the distribution of reported metallicity for these M dwarfs. We find an average metallicity of $-$0.15\,$\pm$\,0.08 for the M dwarfs sample, which is marginally lower than the reported metallicity of $-$0.06\,$\pm$\,0.18 for the FGK primaries from \cite{mann2013prospecting}. The residual star-by-star comparison reveals a difference of 0.14\,$\pm$\,0.09 dex and 0.09\,$\pm$\,0.07 dex, with respect to the metallicity of their FGK primaries from \cite{mann2013prospecting} and the averaged metallicity from the literature, respectively\footnote{The mean and standard deviation of the residual values are calculated after 3-$\sigma$ clipping.}. Note that these values do not indicate any methodological uncertainty and should not be interpreted as suggesting that measurements for M dwarfs have a lower uncertainty than those for FGK stars. They solely represent the dispersion of the determined values for this small sample of stars.  As shown in  Table~\ref{table:elements}, abundance uncertainties of M dwarfs inferred from larger samples are in general similar to those inferred for FGK stars.

Our results support a similar metallicity between M dwarfs and their FGK counterparts within an uncertainty of $\sim$0.1 dex. As shown in Figure \ref{fig:metallicity_binaries}, HD\,154363B may be an exception. While more observations are needed to determine whether such an anomaly is genuine or not, this comparison exercise strongly suggests that elemental abundances derived from NIR high-resolution spectroscopy, as per our methodology, are reliable and free of any systematic effects larger than about 0.1 dex.

\subsection{SLS Sample Properties Compared to Other Studies}

\begin{figure*}
    \centering
    \includegraphics[width=0.75\linewidth]{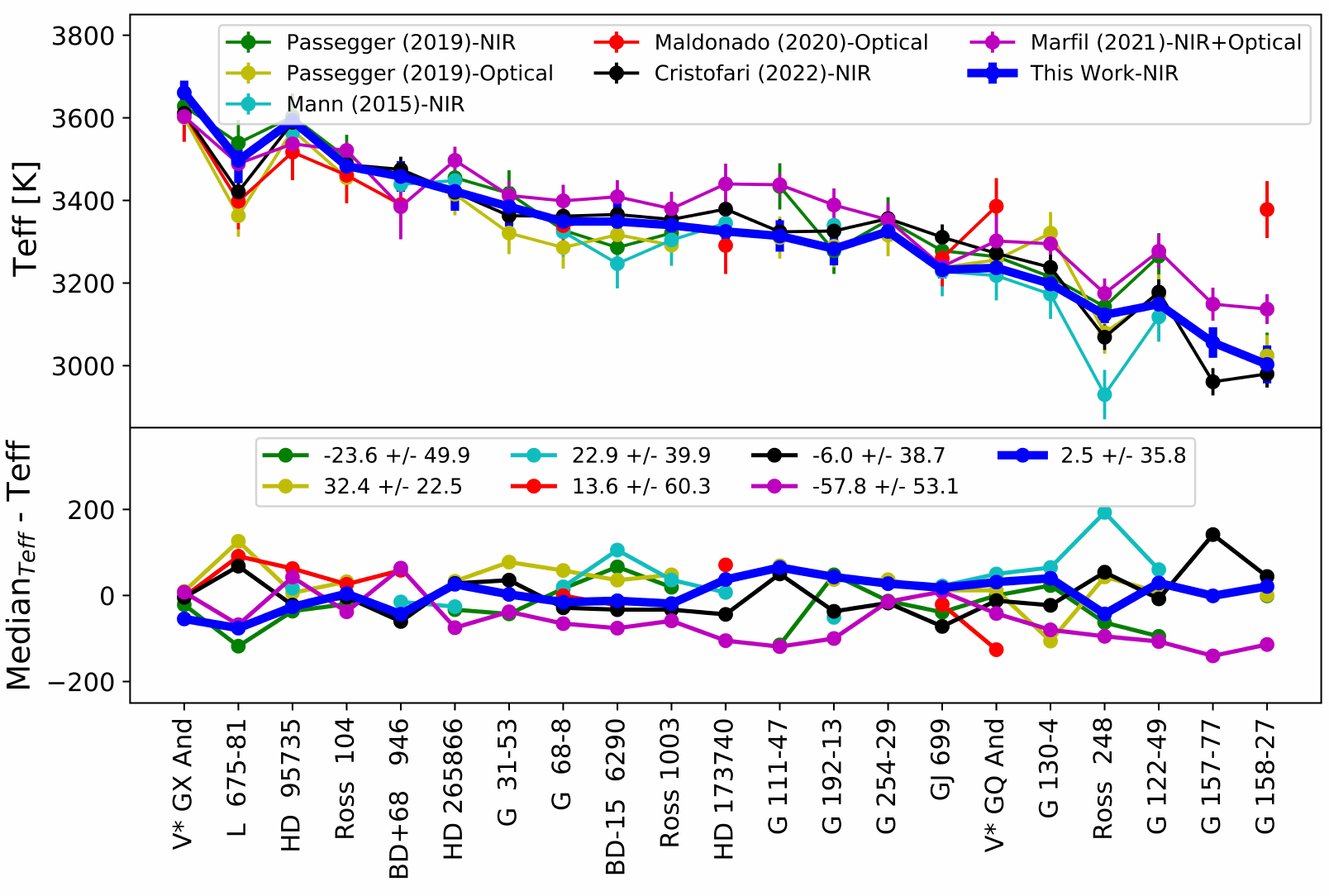}
    \caption{\textit{Top panel}: Comparison of our \teff estimates (blue line) against results of \cite{mann2015constrain}, \cite{passegger2019carmenes}, \cite{maldonado2020hades}, \cite{marfil2021carmenes} and \cite{cristofari2022estimating2}. All our measurements are within 2$\sigma$ of the measurements of most of the other works. \textit{Bottom panel}: Residuals are plotted to compare the median \teff values from all referenced studies against each individual work. The residual measured for \cite{maldonado2020hades} does not include G 158-27 as it was an outlier compared to other measurements of this star.}
    \label{fig:teff_literature}
\end{figure*}
\begin{figure*}
    \centering
    \includegraphics[width=0.75\linewidth]{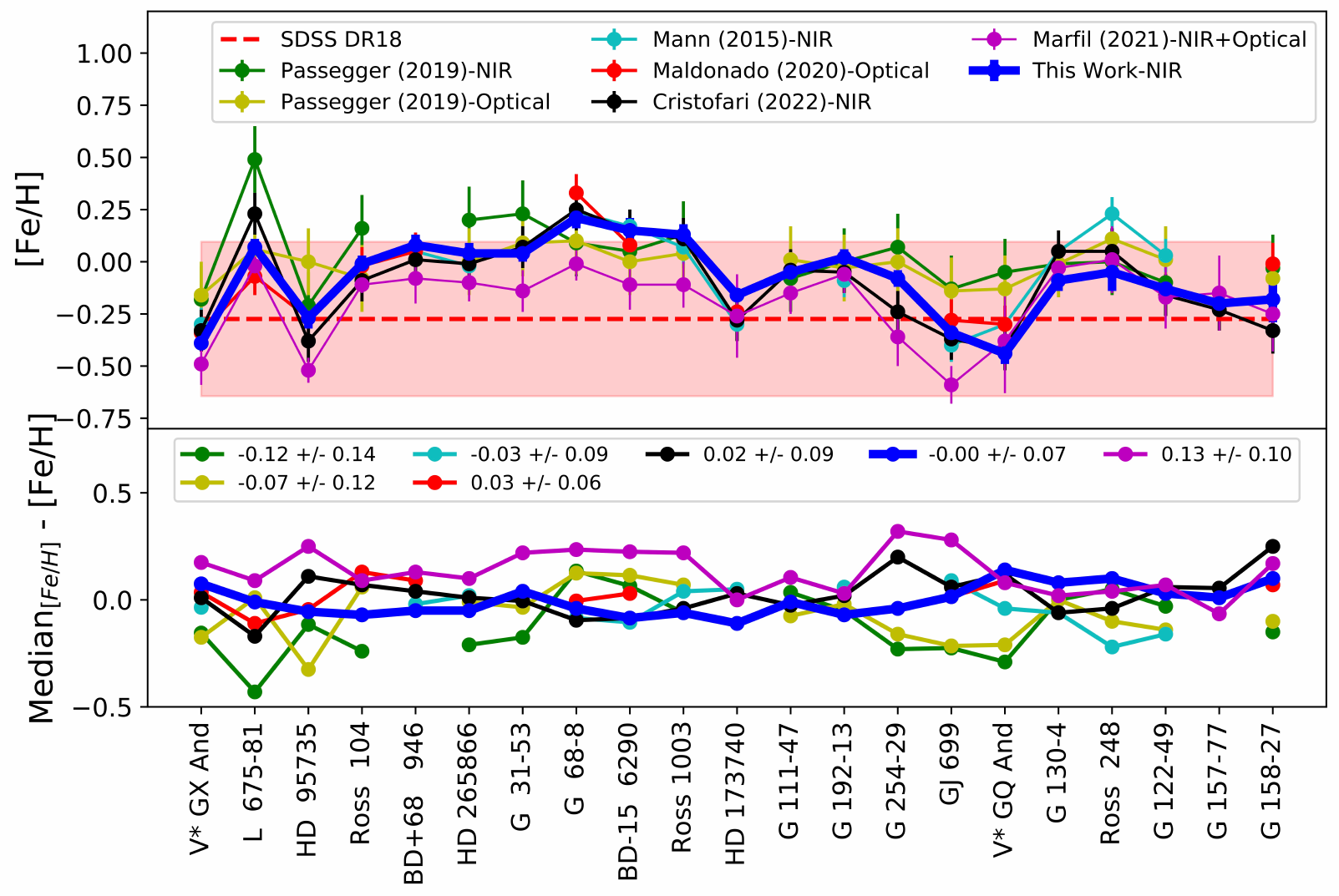}
    \caption{\textit{Top panel}: Comparison of our determined [Fe/H] values (blue line) against the reported metallicity from  \cite{mann2015constrain}, \cite{passegger2019carmenes}, \cite{maldonado2020hades}, \cite{marfil2021carmenes} and \cite{cristofari2022estimating2}. The red shaded area represents the typical distribution of M dwarfs metallicity from SDSS DR18 (\citealt{almeida2023eighteenth}). \textit{Bottom panel}: Residuals are plotted to compare the median metallicity values from all referenced studies against each individual work. All measurements are consistent.}
    \label{fig:met_literature}
\end{figure*}

The comparison exercise conducted on the binary system can be extended to the SLS sample by comparing their \teff and metallicity, both inferred in this study and from other relevant studies in the literature: \cite{mann2015constrain}, \cite{passegger2019carmenes}, \cite{maldonado2020hades}, \cite{marfil2021carmenes}, and \cite{cristofari2022estimating2}. These various samples are described and discussed below.

\cite{mann2015constrain} determined the \teff by comparing optical spectra with BT-Settl PHOENIX models and determined [Fe/H] using a calibration equation from their earlier works (\citealt{mann2013prospecting} and \citealt{mann2014prospecting}), which was calibrated based on several M dwarfs-FGK binary systems. 

\cite{passegger2019carmenes} conducted high-resolution spectroscopy on 282 M dwarfs, several in common with our SLS sample. They utilized PHOENIX models and CARMENES spectra in both visible and NIR wavelengths to determine both \teff and [M/H]. This is especially relevant to our study as it enables a comparison of \teff and abundances derived from optical versus NIR data. 

\cite{maldonado2020hades} developed a unique method incorporating Principal Component Analysis (PCA) to simplify the complexity of spectral data, combined with Sparse Bayesian methods for model fitting to determine fundamental parameters, including [Fe/H]. They calibrated their method using a training set of M dwarfs in binary systems with well-characterized FGK primaries, and then applied it to a few hundred M dwarfs, many of which are in common with the SLS sample. 

\cite{marfil2021carmenes} used high-SNR CARMENES spectra of $\sim$300 M dwarfs to determine [M/H] under different scenarios, such as with free and fixed parameters like \loggns. For consistency, we compare our results with those of \cite{marfil2021carmenes} where \logg is fixed. 

\cite{cristofari2022estimating2} analyzed the same SPIRou data used in this study using a $\chi^2$ minimization method based on MARCS synthetic models (\citealt{gustafsson2008grid}) and a different line list used in our work. They determined various fundamental properties of all our non-binary SLS M dwarfs. This work is particularly important for our analysis as it allows us to directly compare results inferred using different synthetic models and methodologies applied on the same data.

Note that while some of these studies reported [Fe/H] from Fe I lines, others reported [M/H] from atomic lines of various elements. As will be discussed in Section \ref{sec:overall_met_fe_h}, [Fe/H] is not necessarily equivalent to [M/H], particularly for M dwarfs whose spectra are dominated by molecular features. However, for the purpose of comparing metallicities across different studies, we include results from both metrics here.

As shown in Figure \ref{fig:teff_literature}, \teff from all studies show reasonably good agreement with one another. However, comparing these measurements to a ``ground truth" defined as the median of all studies for a given star, one can see significant dispersion among various studies, ranging from 22\,K to 60\,K, ours being among the lowest (35.8\,K) after Passegger et al.\ (\citeyear[22.5\,K]{passegger2019carmenes})
based solely on optical measurements. This is suggestive that our NIR method performs approximately as good as optical ones. 

Similarly for metallicity (see Figure \ref{fig:met_literature}), all studies are in good agreement. We also incorporate a shaded area into Figure \ref{fig:met_literature} to represent the typical distribution of M dwarfs metallicity from SDSS DR18  (\citealt{almeida2023eighteenth}) inferred from a sample of 449 M dwarfs. The SDSS DR18 metallicities appears systematically sub-solar by $\sim$0.3 dex compared to the other studies. This discrepancy will be discussed in the following sections.

\subsection{Elemental Abundances}

\begin{deluxetable*}{crcrc}
\tabletypesize{\normalsize} 
\tablecaption{Mean, standard deviation and relative precision of abundance ratios for our M dwarf sample and other references}
\tablehead{
\colhead{$[$X/H$]$} & \colhead{This Work\,(M)} & \colhead{Maldonado\,(M)} & \colhead{SDSS DR18\,(M)} & \colhead{Brewer\,(FGK)}}

\startdata
    Fe  &	$-$0.11\,$\pm$\,0.16 $[$0.03$]$ &	$-$0.11\,$\pm$\,0.13 $[$0.01$]$ &	$-$0.27\,$\pm$\,0.37  $[$0.02$]$ &	~~\,0.04\,$\pm$\,0.20  $[<$0.01$]$ \\
    Mg &	$-$0.01\,$\pm$\,0.26 $[$0.05$]$ &	$-$0.27\,$\pm$\,0.21 $[$0.01$]$ &	$-$0.26\,$\pm$\,0.26  $[$0.01$]$ &	 ~~\,0.00\,$\pm$\,0.16  $[<$0.01$]$ \\
    Si  &	$-$0.08\,$\pm$\,0.29 $[$0.05$]$ &	$-$0.02\,$\pm$\,0.19 $[$0.01$]$ &	$-$0.27\,$\pm$\,0.32  $[$0.01$]$ &	 ~~\,0.00\,$\pm$\,0.19  $[<$0.01$]$ \\
    Ca &	$-$0.27\,$\pm$\,0.18 $[$0.03$]$ &	$-$0.17\,$\pm$\,0.10 $[$0.01$]$ &	$-$0.22\,$\pm$\,0.32  $[$0.02$]$ &	 ~~\,0.05\,$\pm$\,0.19  $[<$0.01$]$ \\
    Ti &	$-$0.05\,$\pm$\,0.28 $[$0.05$]$ &	$-$0.01\,$\pm$\,0.10 $[$0.01$]$ &	$-$0.36\,$\pm$\,0.35  $[$0.02$]$ &	 ~~\,0.05\,$\pm$\,0.16  $[<$0.01$]$ \\
    Al &	$-$0.10\,$\pm$\,0.26 $[$0.05$]$ &	 ~~0.03\,$\pm$\,0.18 $[$0.01$]$ &	$-$0.29\,$\pm$\,0.71  $[$0.09$]$ &	 ~~\,0.01\,$\pm$\,0.21  $[\sim$0.01$]$  \\
    Na &	$-$0.04\,$\pm$\,0.31 $[$0.05$]$ &	$-$0.04\,$\pm$\,0.16 $[$0.01$]$ &	$-$0.36\,$\pm$\,0.68  $[$0.09$]$ &	$-$0.01\,$\pm$\,0.59  $[\sim$0.01$]$  \\
    C  &	$-$0.02\,$\pm$\,0.28 $[$0.05$]$ &	$-$0.08\,$\pm$\,0.22 $[$0.02$]$ &	$-$0.31\,$\pm$\,0.40  $[$0.02$]$ &	$-$0.02\,$\pm$\,0.18  $[<$0.01$]$ \\
    O  &	$-$0.24\,$\pm$\,0.13 $[$0.03$]$ &	---  &	$-$0.23\,$\pm$\,0.24  $[$0.01$]$ &	~~\,0.08\,$\pm$\,0.17  $[<$0.01$]$ \\
    K  &	$-$0.23\,$\pm$\,0.35 $[$0.07$]$ &	---  &	$-$0.12\,$\pm$\,0.26  $[$0.01$]$ &	--- \\
    % \vspace{0.05cm}
\enddata
\tablecomments{Each value represents the mean\,$\pm$\,standard deviation [relative precision] for each database, where relative precision is defined as standard deviation divided by $\sqrt{N-1}$, where N represents the number of stars in each study.
}
\label{table:elements}
\end{deluxetable*}

We measured the chemical abundances of up to ten different elements, namely: Fe, Mg, O, Si, Ca, Ti, Al, Na, C, and K, for both the binary and the SLS samples (see Table \ref{table:abundances}). Here we compare the distribution of our results for each element with those from three major surveys that provided elemental abundances for various M and FGK samples: SDSS DR18 (M), \cite{maldonado2020hades} (M), and \cite{brewer2016spectral} (FGK). The statistics for all distributions are presented in Table \ref{table:elements}.

The SDSS DR18, the first data release in the fifth phase of the Sloan Digital Sky Survey (SDSS-V), offers mid-level resolution (R = 22\,500) $H$-band spectra of hundreds of thousands of stars, covering various parts of our Galaxy. This database provides elemental abundances for 449 M dwarfs, enabling a bulk examination of nearby M dwarfs' properties.

\cite{brewer2016spectral} conducted an extensive spectroscopic study on 1617 FGK stars, using data from the California Planet Search program obtained with the HIRES spectrograph at Keck Observatory (R\,$\sim$\,70,000). They used Atlas9 (\citealt{castelli2004new}) one-dimensional LTE models in plane-parallel atmospheres for their analysis. Covering a spectral range between 516.4\,-\,780 nm, they determined \teffns, \loggns, metallicity, rotational velocity, and the abundances of 15 elements.

\citet{maldonado2020hades} used PCA and Sparse Bayesian techniques to analyze the spectral data of M dwarfs, focusing on determining their fundamental properties and chemical abundances. The method was then calibrated using M dwarfs in binary systems with FGK primaries and applied to tens of M dwarfs.

As shown in Table \ref{table:elements}, Figure  \ref{fig:master_chemistry} and \ref{sec:p2_appendix2} (in appendix), our chemical distributions align well within standard deviations of the mentioned surveys. Our results for all elements, except perhaps for Ca and O, are consistent with the solar values, considering their large standard deviation. Notably, our [O/H] and [Ca/H] are marginally sub-solar which is intriguing. Given the high-precision of our oxygen abundance measurements from several OH lines, compared to other elements, this suggests three possibilities: an unusual oxygen depletion in our M dwarf sample, a systematic underestimation of OH lines in the synthetic models or some unknown flaws in our methodology. The latter appears unlikely given our reliable \teff measurements presented above involving a methodology very similar to that used for elemental abundance determination.  

\begin{figure*}
    \centering
    \includegraphics[width=0.75\linewidth]{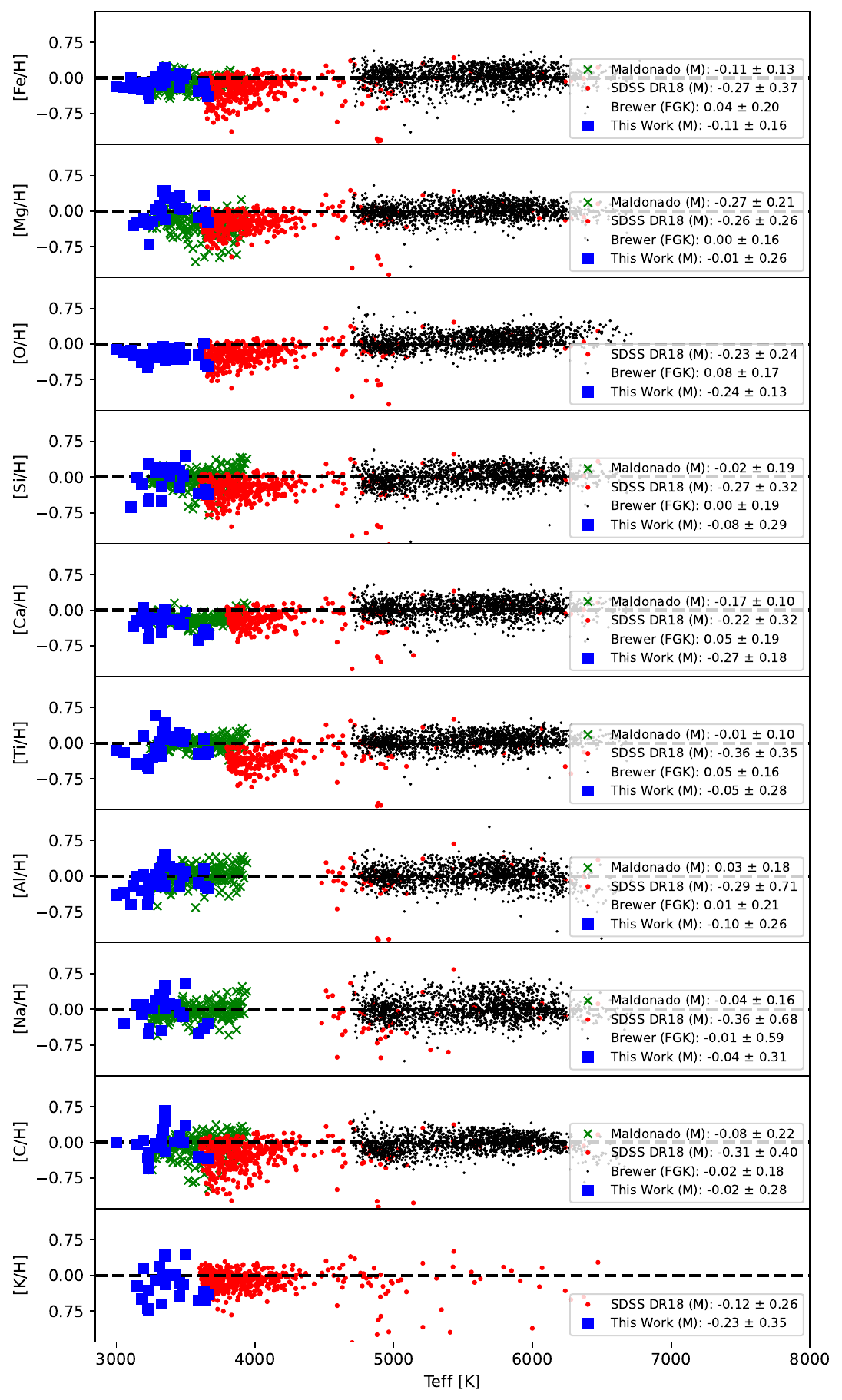}
    \caption{Comparative analysis of the chemical abundances in M dwarfs, from various datasets: our study, the SDSS DR18 database and \cite{maldonado2020hades}. Each subplot represents a different chemical element, showing the relative abundances across a range of temperatures. For a broader contextual understanding, we have also included the chemical abundances of FGK stars from the study by \cite{brewer2016spectral}, represented as black dots. Note that uncertainties in all studies are typically less than 0.1\,dex.
}
    \label{fig:master_chemistry}
\end{figure*}

All the SDSS DR18 abundances for M dwarfs tend to be systematically lower than those for FGK stars. As discussed in \cite{Jahandar2023}, isolating some atomic lines from OH bands in high-resolution spectra is challenging, especially for cold targets. An instance of this is the disappearance of the Si line at 1589.27\,nm in sub-solar metallicity stars cooler than 3600\,K (see Figure 16 in \citealt{Jahandar2023}). Given the relatively low spectral resolution ($\sim$22\,500) of the SDSS database and the cold \teff of M dwarfs, the striking similarity between all elemental distributions from the SDSS database may be suggestive that some measured lines are contaminated by molecular features such as OH lines that are very numerous in the NIR. However, since OH lines are so numerous in the NIR, their abundance is very well constrained. Thus, the relatively low [O/H] abundance from SDSS and our work (i.e. $-$0.24$\,\pm\,$0.03) compared to FGK stars (0.08\,$\pm$\,0.01), a 10-$\sigma$ difference, may be suggestive of an unusual depletion of oxygen in M dwarfs compared to FGK stars. However, with no obvious mechanism for such depletion, these results remain highly speculative.

Finally, it is worth noting that the mean magnesium abundances from \cite{maldonado2020hades} are also notably lower compared to our findings, the FGK stars, and their other measurements. The impact of this will be discussed in Section \ref{sec:refractory}.

\subsection{Overall Metallicity vs [Fe/H]} \label{sec:overall_met_fe_h}

One common metric for quantifying metallicity in stars is the ratio of iron to hydrogen, denoted as [Fe/H]. Iron is preferred as a metallicity indicator partially due to the presence of numerous strong spectral lines that are relatively easy to observe and measure. In many studies, the term ``Overall Metallicity" and [Fe/H] are often used interchangeably. This usage stems from the assumption that the abundance ratios of all elements scale similarly with iron. However, this simplifying assumption may not always be accurate, as abundance ratios can vary significantly among stars with similar [Fe/H] values.

Historically, the lack of high-resolution spectroscopy often necessitated the use of overall metallicity as a substitute for [Fe/H]. Typically, overall metallicity was inferred from Spectral Energy Distribution (SED) fitting of the entire continuum, or by using metal-sensitive spectral features (\citealt{Stassun_2016}; \citealt{Stassun_2017}; \citealt{cadieux2022toi}; \citealt{gan2023massive}). These methods were particularly useful given technological or observational constraints. However, advancements in spectroscopic techniques have revealed that the relationship between [Fe/H] and overall metallicity is more complex. This understanding necessitates careful consideration when using [Fe/H] as a proxy for overall metallicity.

The distinction between [Fe/H] and overall metallicity is crucial for the interior modeling of exoplanets. Relying on overall metallicity could potentially skew results due to the influence of non-Fe lines. This issue is particularly pronounced in M dwarfs dominated by molecular bands like OH, where overall metallicity might align more with the oxygen abundance than with [Fe/H]. Consequently, a more precise measurement of [Fe/H] using Fe lines is preferred over using overall metallicity as a proxy.

\begin{figure}[ht]
    \centering
    \includegraphics[width=1\linewidth]{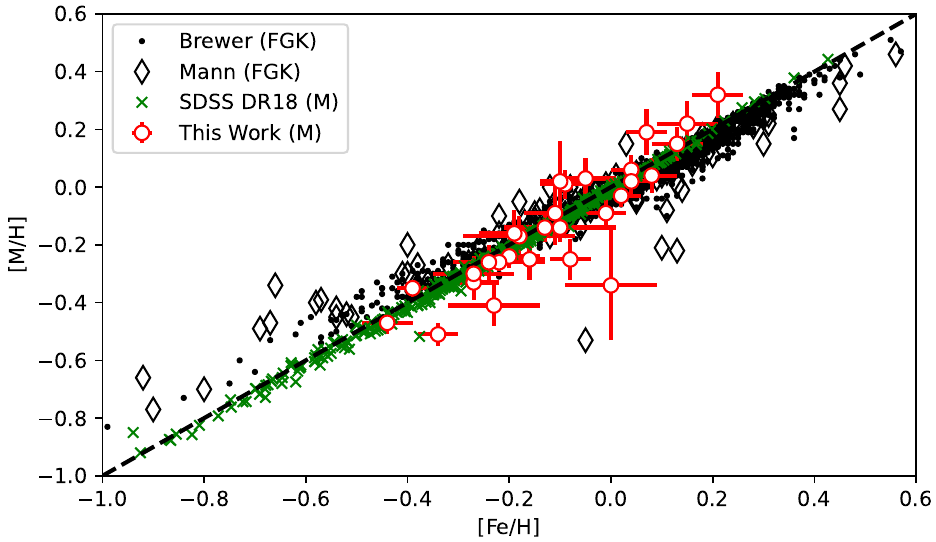}
    \caption{Overall metallicity, [M/H], correlates with [Fe/H] in our M dwarf sample, SDSS DR18 M dwarfs, and FGK stars from the studies by \cite{mann2013prospecting} and \cite{brewer2016spectral}. While there is a consistent correlation, not all stars, show similar values. Overall metallicity determination, unlike line-by-line analysis, accounts for all elements. For FGK stars, the impact is minimal, but for M dwarfs, which often have spectra dominated by OH or H$_2$O lines, the overall metallicity (such as the metallicity derived from SED fitting) is more influenced by oxygen levels. Consequently, for stars with depleted oxygen, there can be a significant discrepancy between [M/H] and [Fe/H]. }
    \label{fig:metallicity_fe_h}
\end{figure}

We examine the differences between overall metallicity and [Fe/H] by comparing these parameters from various studies, including our work (M), \cite{brewer2016spectral} (FGK), \cite{mann2013spectro}(FGK) and the SDSS DR18 database (M) in Figure \ref{fig:metallicity_fe_h}. While a correlation between [Fe/H] and [M/H] is evident, there are instances where the overall metallicity significantly differs from [Fe/H].
With the advent of NIR and optical high-resolution spectroscopy, [M/H] is an ill-defined quantity likely wavelength-dependent since some lines, like OH in the NIR, can contaminate other elements. Since OH dominates [M/H] in the NIR and can contaminate Fe lines, it may explain the relatively tight correlation observed in the SDSS data (see Figure \ref{fig:metallicity_fe_h}) based on medium resolution spectroscopy. Contamination from OH lines may also explain the small dispersion observed in the SDSS elemental abundances (see Table \ref{table:elements}). In this context, [Fe/H] is a much better and unambiguous metallicity indicator that should be preferred over [M/H]. This comparative analysis emphasizes the need for high resolution spectroscopy for precise unbiased chemical abundance determination from NIR data.

\subsection{Refractory Elements}\label{sec:refractory}

\begin{deluxetable*}{cccccc}[ht]
\tablecaption{Distribution of Molar Abundance Ratios}
\tabletypesize{\normalsize} 
\tablehead{
\colhead{X/H} & \colhead{This Work (M)} & \colhead{Sun$^{\dag}$} & SDSS DR18 (M)& Maldonado (M) & Brewer (FGK)}

\startdata
C/O   & 0.82 $\pm$ 0.21  & 0.55\,$\pm$\,0.17  & 0.50 $\pm$ 0.13    & ---               &  0.45 $\pm$ 0.10      \\
Mg/Si & 1.62 $\pm$ 0.74  & 1.23\,$\pm$\,0.12  & 1.20 $\pm$ 0.13    & 0.73 $\pm$ 0.24   &  1.25 $\pm$ 0.18      \\
Fe/Mg & 0.69 $\pm$ 0.26  & 0.79\,$\pm$\,0.14  & 0.84 $\pm$ 0.18    & 1.14 $\pm$ 0.38   &  0.86 $\pm$ 0.13      \\
Fe/O  & 0.09 $\pm$ 0.03  & 0.07\,$\pm$\,0.02  & 0.06 $\pm$ 0.01    & ---               &  0.06 $\pm$ 0.02    \\
\enddata
\tablecomments{
$^{\dag}$ Photospheric abundance ratios from \citet{asplund2009chemical}.\\
For all the reported estimates, except for the Sun and FGK stars, they represent the general distribution of M dwarfs in various studies, after applying 3 sigma clipping.}

\label{table:ratios}
\end{deluxetable*}

\begin{figure*}[ht]
    \centering
    \includegraphics[width=0.7\linewidth]{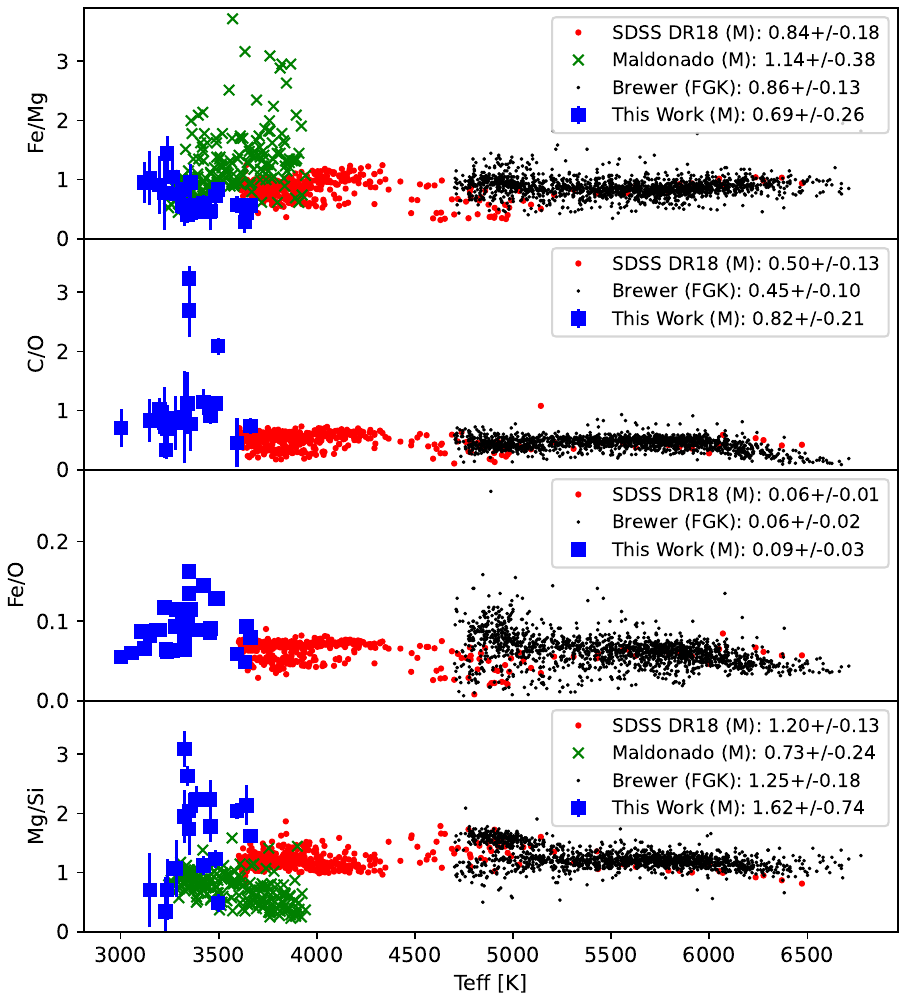}
    \caption{Comparative analysis of four different molar ratios in M dwarfs from three distinct datasets: our study, the SDSS DR18 database, and \cite{maldonado2020hades}. Each subplot represents a different chemical element, showing the relative ratio across a range of temperatures. For a broader contextual understanding, we have also included the chemical abundances of FGK stars from the study by \cite{brewer2016spectral}, represented as black dots. The values reported in the legend are after 3 sigma clipping. The typical uncertainty in our Fe/O measurements is $\sim$\,0.15, which is not shown in the figure to avoid overcrowding.}
    \label{fig:master_ratio_chemistry}
\end{figure*}

Refractory elements (Fe, Mg, Si) are critical for the interior modeling of exoplanets, significantly contributing to the bulk composition of planets and influencing their internal structure and thermal evolution. These elements, alongside the C/O ratio, determine the chemical and physical properties of the planets and their atmospheres. The C/O ratio, in particular, offers insights into the formation location of gas giant exoplanets within the circumstellar disk (\citealt{Oberg2011}; \citealt{favre2013significantly}; \citealt{madhusudhan2014toward}). The abundance of refractory elements, combined with the C/O ratio in giant planets and their host stars, is crucial for understanding their formation histories and the conditions within the protoplanetary disk where they originated (\citealt{wang2019enhanced}; \citealt{lothringer2021new}).

Table \ref{table:ratios} and Figure \ref{fig:master_ratio_chemistry} present the C/O, Mg/Si, Fe/Mg, and Fe/O ratios of our M dwarfs, the Sun, FGK stars from \cite{brewer2016spectral}, M dwarfs from SDSS DR18, and \cite{maldonado2020hades}. While the abundance ratios are generally consistent with the Sun and other studies some significant dispersion is observed, e.g. our C/O which appears slightly higher compared to that from SDSS (M) and \cite{brewer2016spectral} (FGK stars). We note also that Mg/Si and Fe/Mg ratios from \cite{maldonado2020hades} are somewhat discrepant compared to other works which may result from lower [Mg/H] values, as mentioned earlier (see also Figure~\ref{fig:master_ratio_chemistry}). The Fe/O ratio displays very small dispersion with excellent agreement across this work, the Sun, SDSS DR18 (M), and \cite{brewer2016spectral} (FGK). 

In Figure \ref{fig:master_ratio_chemistry}, we illustrate the scatter of molar ratios for all stars in each study. Our Fe/Mg measurements are in agreement with FGK and SDSS DR18 data. In contrast, the data from \cite{maldonado2020hades} tends to scatter upward, indicating a possible overestimation of the Fe/Mg ratio, as a result of the lower [Mg/H] in their measurements highlighted earlier. As noted before, the C/O ratios in most of our stars are slightly higher than those of FGK stars and SDSS DR18 M dwarfs. Specifically, three stars show a significantly high C/O, which, if not a result of caveats in our synthetic models or methodology, may suggest stars with unusually high C/O ratios, requiring further investigation. The Fe/O ratios are in excellent agreement in all studies, considering the narrow range of the y-axis in the third panel of Figure \ref{fig:master_ratio_chemistry}. Nevertheless, the distribution of our Fe/O data closely mirrors that found in colder FGK stars. Finally, the Mg/Si ratios in FGK stars show an unexpected bimodal distribution, and our data suggest a continuation of this trend into cooler stellar temperatures. The systematic lower Mg/Si values from \cite{maldonado2020hades}, when compared with our own and those from FGK stars, reiterate the influence of their low [Mg/H] estimates.

\begin{figure*}
    \centering
    \includegraphics[width=0.8\linewidth]{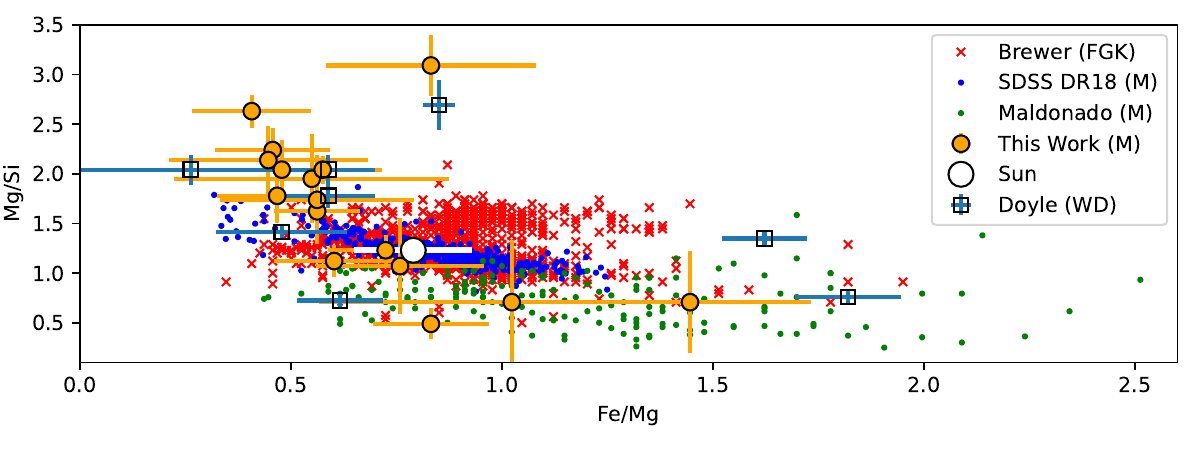}
    \caption{Mg/Si vs. Fe/Mg ratios for various stellar populations. The figure features FGK stars (red crosses), SDSS DR18 data (blue dots), M dwarfs from \cite{maldonado2020hades} (green dots), and results from this work (orange circles), unveiling a non-linear correlation between these ratios. Additionally, white dwarfs (\citealt{doyle2023new}) are included for a more comprehensive comparison. The FGK stars exhibit two patterns: one follows the non-linear correlation consistent with the other studies, while the other is nearly flat.}
    \label{fig:mg_si_fe_mg_v2}
\end{figure*}

Additionally, Figure \ref{fig:mg_si_fe_mg_v2} illustrates the relationship between the Mg/Si and Fe/Mg ratios across different stellar classifications, including data from oxygen-bearing White Dwarfs (WDs, \citealt{doyle2023new}). 
FGK stars display a bimodal distribution that appears to be very different from the SDSS M dwarfs sample which is somewhat intriguing.  For WDs, the refractory elements reflect the abundance of asteroid debris illustrating that circumstellar disk material can take a fairly wide range of refractory elemental ratio. The wide dispersion observed from the SLS sample appears to be consistent with that from WDs, suggesting that the observed large dispersion is probably real. This figure further illustrates that the ratios from \cite{maldonado2020hades} are somewhat discrepant when compared to those from the M (SDSS) and FGK samples. Except for one case showing a relatively high Mg/Si for its measured Fe/Mg, the ratios from this work are consistent with the broad distributions observed in different studies.

\section{Conclusion} \label{sec:conclusion}

This work, as an extension to \cite{Jahandar2023}, represents a comprehensive high-resolution infrared spectroscopy study of 31 M dwarfs in the solar neighborhood, including 10 M dwarfs in binary systems with FGK primaries.

We analyzed the high-resolution spectra of our target stars from the SPIRou instrument using $\chi^2$ analysis on hundreds of absorption lines, comparing them to PHOENIX-ACES synthetic spectra. This analysis enabled us to determine the \teffns, overall metallicity, and chemical abundances of up to 10 different elements. We calibrated our \teff method by testing it against synthetic models at various signal-to-noise ratios, finding a consistent inherent uncertainty of 10\,K for high-SNR observations. Empirically, the uncertainty of \teff in this work is found to be $\sim$30\,K for stars with SNR$>$100. Our \teff values correlate well with interferometric measurements, with a typical standard deviation of $\sim$30\,K in both methods. This provides a validation of our methodology for deriving \teff through NIR high-resolution spectroscopy.

Using our \teff values and \logg from the TESS database, along with our refined line list from the NIST and BT-Settl PHOENIX databases, we measured the chemical abundances of ten elements, including refractory elements like Si, Mg, and Fe, crucial for exoplanet interior modeling. Our results aligned with several recent independent studies. We found an average metallicity of $-$0.11\,$\pm$\,0.16 for our targeted M dwarfs, consistent with the findings of \cite{maldonado2020hades}, SDSS DR18, and FGK stars from \cite{brewer2016spectral}, considering the standard deviations in all studies. Similarly, in binary systems, our findings indicated an average metallicity of $-$0.15\,$\pm$\,0.08 for M dwarfs, which was marginally lower yet consistent with the $-$0.06\,$\pm$\,0.18 for the FGK primaries reported by \cite{mann2013prospecting}. 

We compared the molar ratios of key refractory elements in our M dwarf sample to those of the Sun, FGK stars, and other M dwarfs from SDSS DR18 and \cite{maldonado2020hades}. Our C/O estimates were slightly higher, while the Fe/Mg, Fe/O, and Mg/Si ratios matched those of cooler FGK stars. We observed that the elemental abundances estimated by SDSS DR18 are likely systematically lower compared to other studies. This underestimation may be attributed to the difficulty in isolating atomic lines within the spectra of cool stars at low and medium resolutions, potentially leading to contamination from OH lines. This study highlights significant discrepancies across major M dwarf surveys, likely resulting from a variety of analysis methodologies involving different atmosphere models. 

Following the procedure explained in \cite{Jahandar2023}, we addressed numerous flaws in synthetic models, such as continuum mismatch, shifted lines, and line contamination. Consequently, this necessitated the exclusion of thousands of spectral features from our observed data. Such comprehensive high-resolution spectroscopy is instrumental in providing detailed feedback, crucial for the refinement of synthetic models.

\section{Acknowledgments}
\par We would like to thank the anonymous referee for the constructive comments and thorough review of this manuscript.
\par This work is based on observations obtained at the Canada-France-Hawaii Telescope (CFHT) which is operated from the summit of Maunakea by the National Research Council of Canada, the Institut National des Sciences de l'Univers of the Centre National de la Recherche Scientifique of France, and the University of Hawaii. The observations at the Canada-France-Hawaii Telescope were performed with care and respect from the summit of Maunakea which is a significant cultural and historic site.
\par This work is partly funded through the National Science and Engineering Research Council of Canada through the Discovery Grant program and the CREATE training program on New Technologies for Canadian Observatories. We also acknowledge the generous financial support of the Trottier Family Foundation for the Trottier Institute for Research on Exoplanets. 
\par We acknowledge funds from the European Research Council (ERC) under the H2020 research \& innovation program (grant
agreement \#740651 NewWorlds)
\par J.H.C.M. is supported by FCT - Fundação para a Ciência e a Tecnologia through national funds by these grants: UIDB/04434/2020, UIDP/04434/2020, PTDC/FIS-AST/4862/2020. J.H.C.M. is also supported by the European Union (ERC, FIERCE, 101052347). Views and opinions expressed are however those of the author(s) only and do not necessarily reflect those of the European Union or the European Research Council. Neither the European Union nor the granting authority can be held responsible for them.


\vspace{5mm}
% \facilities{HST(STIS), Swift(XRT and UVOT), AAVSO, CTIO:1.3m,


\software{\texttt{Astropy} \citep{Astropy_2018}; \texttt{matplotlib} \citep{Hunter_2007}; \texttt{SciPy} \citep{Virtanen_2020}; \texttt{NumPy} \citep{Harris_2020}}.



\pagebreak

\bibliography{sample631}{}
\bibliographystyle{aasjournal}

\appendix
\counterwithin{table}{section}
\counterwithin{figure}{section}
\vspace{1cm}
\section{Detection Criteria of Elemental Spectral Lines} \label{sec:appendix2}

Due to the molecular-dominated nature of M dwarfs, some spectral lines are only visible at certain \teff and metallicities. In Table \ref{table:line_critera}, we list each spectral line used in this analysis that has been visible in at least five different M dwarfs. We also add six columns for the minimum and maximum \teffns, [M/H] and [Fe/H] of the stars that had this line in their spectra. The \teff and metallicity values are rounded to the nearest 100\,K and 0.1\, dex, respectively. Note that clear observation of these lines is also subject to the high SNR of the data. 
\vspace{0.1cm}

\begin{longtable}{clccccccc}
\caption{Detection Criteria of Elemental Spectral Lines}\label{table:line_critera}\\
\hline
\colhead{Element} & \colhead{Wavelength} & \colhead{Min \teff} & \colhead{Max \teff} & \colhead{Min [M/H]} & \colhead{Max [M/H]} & \colhead{Min [Fe/H]} & \colhead{Max [Fe/H]} & \colhead{Ref}\\
\noalign{\vskip -7pt} % Adjust the negative space as needed
\colhead{} & \colhead{(vacuum, nm)} & \colhead{(K)} & \colhead{(K)} & \colhead{(dex)} & \colhead{(dex)} & \colhead{(dex)} & \colhead{(dex)} & \colhead{} \\
\hline
\endhead
\hline
\endfoot
    Fe I &         1014.835 &      3500 &      3700 &      $-$0.2 &        0.1 &       $-$0.4 &         0.1 &  BT-Settl  \\
    Fe I &         1015.797 &      3400 &      3700 &      $-$0.2 &        0.1 &       $-$0.4 &         0.1 &  BT-Settl  \\
    Fe I &         1017.027 &      3300 &      3700 &      $-$0.2 &        0.2 &       $-$0.4 &         0.2 &  BT-Settl  \\
    Fe I &         1034.373 &      3400 &      3700 &      $-$0.2 &        0.1 &       $-$0.4 &         0.0 &  BT-Settl  \\
    Fe I &         1042.586 &      3300 &      3700 &      $-$0.2 &        0.1 &       $-$0.4 &         0.1 &  BT-Settl  \\
    Fe I &         1112.284&      3300 &      3700 &      $-$0.2 &       0.2 &       $-$0.4 &        0.2 &  BT-Settl  \\
   Fe I &         1125.421 &      3100 &      3700 &      $-$0.2 &        0.2 &       $-$0.4 &         0.2 &  NIST  \\
   Fe I &         1137.719 &      3200 &      3700 &      $-$0.2 &        0.2 &       $-$0.4 &         0.2 &  NIST  \\
   Fe I &         1139.405 &      3200 &      3700 &      $-$0.3 &        0.2 &       $-$0.4 &         0.2 &  NIST  \\
   Fe I &         1144.227 &      3200 &      3700 &      $-$0.2 &        0.1 &       $-$0.4 &         0.1 &  NIST  \\
    Fe I &         1159.677 &      3400 &      3600 &     $-$0.0 &        0.1 &       $-$0.0 &         0.1 &  BT-Settl  \\
   Fe I &         1256.044 &      3100 &      3600 &      $-$0.1 &        0.1 &       $-$0.3 &         0.2 &  NIST  \\
    Fe I &         1281.063 &      3300 &      3700 &     $-$0.2 &        0.2 &       $-$0.4 &         0.2 &  BT-Settl  \\
    Fe I &         1288.331 &      3500 &      3600 &     $-$0.2 &        0.0 &       $-$0.3 &      $-$0.0 &  NIST  \\
   Fe I &         1360.436 &      3200 &      3500 &      $-$0.3 &        0.2 &        $-$0.4 &         0.2 &  NIST  \\
   Fe I &         1372.872 &      3000 &      3200 &      $-$0.2 &     $-$0.0 &       $-$0.4 &      $-$0.1 &  NIST  \\
   Fe I &         1420.613 &      3100 &      3300 &      $-$0.0 &        0.0 &       $-$0.1 &         0.0 &  NIST  \\
   Fe I &         1463.210 &      3200 &      3700 &      $-$0.3 &        0.2 &       $-$0.4 &         0.2 &  NIST  \\
   Fe I &         1477.156 &      3300 &      3600 &         0.0 &        0.2 &       $-$0.0 &         0.2 &  NIST  \\
   Fe I &         1484.987 &      3200 &      3600 &      $-$0.2 &        0.1 &       $-$0.3 &         0.2 &  NIST  \\
   Fe I &         1550.503 &      3200 &      3600 &      $-$0.2 &        0.2 &       $-$0.4 &         0.2 &  NIST  \\
   Fe I &         1562.543 &      3100 &      3600 &      $-$0.1 &        0.1 &       $-$0.2 &         0.1 &  NIST  \\
   Fe I &         1562.595 &      3000 &      3700 &      $-$0.2 &        0.2 &       $-$0.4 &         0.2 &  BT-Settl  \\
   Fe I &         1572.787 &      3400 &      3600 &      $-$0.0 &        0.1 &       $-$0.2 &         0.0 &  NIST  \\
   Fe I &         1575.349$^*$&      3100 &      3700 &      $-$0.2 &        0.2 &       $-$0.4 &         0.2 &  NIST  \\
   Fe I &         1578.105 &      3500 &      3600 &      $-$0.2 &        0.0 &       $-$0.3 &      $-$0.0 &  NIST  \\
   Fe I &         1583.362 &      3300 &      3600 &         0.0 &        0.2 &          0.0 &         0.2 &  NIST  \\
   Fe I &         1604.710 &      3300 &      3600 &      $-$0.1 &        0.1 &       $-$0.3 &         0.0 &  NIST  \\
   Fe I &         1641.486 &      3200 &      3500 &      $-$0.2 &        0.0 &       $-$0.4 &      $-$0.1 &  NIST  \\
   Fe I &         1644.934 &      3300 &      3700 &      $-$0.2 &        0.2 &       $-$0.4 &         0.2 &  NIST  \\
   Fe I &         1661.217$^*$&      3100 &      3600 &      $-$0.3 &        0.0 &       $-$0.4 &         0.0 &  NIST  \\
   Fe I &         1672.364$^*$&      3100 &      3700 &      $-$0.3 &        0.2 &       $-$0.4 &         0.2 &  NIST  \\
   Fe I &         1675.373$^*$&      3100 &      3700 &      $-$0.3 &        0.2 &       $-$0.4 &         0.2 &  NIST  \\
   Fe I &         1675.765 &      3000 &      3400 &      $-$0.2 &        0.2 &       $-$0.4 &         0.2 &  NIST  \\
   Fe I &         1688.941$^*$&      3100 &      3400 &      $-$0.3 &        0.1 &       $-$0.4 &         0.1 &  NIST  \\
   Fe I &         1690.350$^*$&      3100 &      3600 &      $-$0.3 &        0.1 &       $-$0.4 &         0.0 &  NIST  \\
   Fe I &         1705.682$^*$&      3000 &      3500 &      $-$0.2 &        0.0 &       $-$0.4 &       $-$0.0 &  NIST  \\
   Fe I &         1733.883$^*$&      3300 &      3700 &      $-$0.2 &        0.1 &       $-$0.4 &         0.1 &  NIST  \\
   Fe I &         1820.395 &      3000 &      3600 &      $-$0.1 &     $-$0.1 &      $-$0.3 &      $-$0.1 &  NIST  \\
   Fe I &         1899.219 &      3300 &      3600 &      $-$0.1 &        0.2 &       $-$0.3 &         0.2 &  NIST  \\
   Mg I &         1081.404 &      3300 &      3700 &      $-$0.2 &        0.2 &       $-$0.4 &         0.2 &  BT-Settl  \\
   Mg I &         1502.911 &      3200 &      3700 &      $-$0.2 &        0.1 &       $-$0.4 &         0.1 &  NIST  \\
   Mg I &         1505.184 &      3200 &      3700 &      $-$0.2 &        0.2 &       $-$0.4 &         0.2 &  BT-Settl  \\
   Mg I &         1574.503 &      3400 &      3700 &      $-$0.2 &        0.2 &       $-$0.4 &         0.2 &  BT-Settl  \\
   Mg I &         1575.328$^*$&      3200 &      3700 &      $-$0.2 &        0.2 &       $-$0.4 &         0.2 &  BT-Settl  \\
   Mg I &         1577.016 &      3300 &      3700 &      $-$0.2 &        0.2 &       $-$0.4 &         0.2 &  BT-Settl  \\
   Mg I &         1711.336 &      3100 &      3700 &      $-$0.2 &        0.2 &       $-$0.4 &         0.2 &  NIST  \\
   Mg I &         2385.059 &      3300 &      3700 &      $-$0.2 &        0.1 &       $-$0.4 &         0.1 &  NIST  \\
   Si I &         1075.231 &      3100 &      3600 &      $-$0.1 &        0.1 &       $-$0.3 &         0.1 &  BT-Settl  \\
   Si I &         1083.006 &      3300 &      3700 &      $-$0.2 &        0.2 &       $-$0.4 &         0.2 &  NIST  \\
   Si I &         1198.748 &      3300 &      3700 &      $-$0.2 &        0.2 &       $-$0.4 &         0.2 &  BT-Settl  \\
   Si I &         2325.604 &      3100 &      3700 &      $-$0.3 &        0.1 &       $-$0.4 &         0.1 &  NIST  \\
   Ca I &         1281.957 &      3100 &      3700 &      $-$0.3 &        0.2 &       $-$0.4 &         0.2 &  NIST  \\
   Ca I &         1282.739 &      3100 &      3700 &      $-$0.2 &        0.2 &       $-$0.4 &         0.2 &  NIST  \\
   Ca I &         1283.056 &      3100 &      3700 &      $-$0.2 &        0.2 &       $-$0.4 &         0.2 &  NIST  \\
   Ca I &         1303.712 &      3200 &      3600 &      $-$0.2 &     $-$0.0 &       $-$0.4 &      $-$0.1 &  NIST  \\
   Ti I &          983.482 &      3000 &      3700 &      $-$0.2 &        0.2 &       $-$0.4 &         0.2 &  BT-Settl  \\
   Ti I &          993.005 &      3300 &      3700 &      $-$0.2 &        0.2 &       $-$0.4 &         0.2 &  BT-Settl  \\
   Ti I &         1000.583 &      3400 &      3700 &      $-$0.2 &        0.1 &       $-$0.4 &         0.1 &  NIST  \\
   Ti I &         1005.159 &      3300 &      3700 &      $-$0.2 &        0.1 &       $-$0.4 &         0.1 &  NIST  \\
   Ti I &         1006.048 &      3300 &      3700 &      $-$0.2 &        0.1 &       $-$0.4 &         0.2 &  NIST  \\
   Ti I &         1012.367 &      3300 &      3700 &      $-$0.2 &        0.2 &       $-$0.4 &         0.2 &  NIST  \\
   Ti I &         1055.586 &      3300 &      3700 &      $-$0.2 &        0.1 &       $-$0.4 &         0.1 &  BT-Settl  \\
   Ti I &         1078.432 &      3400 &      3500 &         0.0 &        0.1 &         0.0 &         0.1 &  NIST  \\
   Ti I &         1081.995 &      3300 &      3600 &      $-$0.0 &        0.2 &       $-$0.0 &         0.2 &  NIST  \\
   Ti I &         1083.639 &      3300 &      3500 &      $-$0.0 &        0.2 &       $-$0.0 &         0.2 &  NIST  \\
   Ti I &         1123.403 &      3300 &      3500 &      $-$0.1 &        0.2 &       $-$0.2 &         0.2 &  NIST  \\
   Ti I &         1178.378 &      3200 &      3700 &      $-$0.3 &        0.1 &       $-$0.4 &         0.1 &  NIST  \\
   Ti I &         1201.923 &      3400 &      3600 &         0.0 &        0.1 &         0.0 &         0.1 &  BT-Settl  \\
   Ti I &         1260.371 &      3200 &      3700 &      $-$0.3 &        0.1 &       $-$0.4 &         0.2 &  NIST  \\
   Ti I &         1267.458 &      3200 &      3700 &      $-$0.3 &        0.1 &       $-$0.4 &         0.1 &  NIST  \\
   Ti I &         1274.840 &      3100 &      3700 &      $-$0.2 &        0.2 &       $-$0.4 &         0.2 &  NIST  \\
   Ti I &         1281.499 &      3000 &      3700 &      $-$0.3 &        0.2 &       $-$0.4 &         0.2 &  BT-Settl  \\
   Ti I &         1292.342 &      3100 &      3700 &      $-$0.3 &        0.2 &       $-$0.4 &         0.2 &  NIST  \\
   Ti I &         1308.086 &      3400 &      3700 &      $-$0.2 &        0.2 &       $-$0.4 &         0.2 &  NIST  \\
   Ti I &         1340.025 &      3200 &      3700 &      $-$0.2 &        0.1 &       $-$0.4 &         0.1 &  NIST  \\
   Ti I &         1345.471 &      3300 &      3600 &      $-$0.1 &        0.1 &       $-$0.3 &         0.1 &  BT-Settl  \\
   Ti I &         1533.902 &      3200 &      3700 &      $-$0.2 &        0.2 &       $-$0.4 &         0.2 &  BT-Settl  \\
   Ti I &         1554.801 &      3300 &      3700 &      $-$0.2 &        0.1 &       $-$0.4 &         0.1 &  BT-Settl  \\
   Ti I &         1571.989 &      3200 &      3700 &      $-$0.2 &        0.2 &       $-$0.4 &         0.2 &  BT-Settl  \\
   Ti I &         1589.638 &      3100 &      3400 &      $-$0.3 &        0.0 &       $-$0.3 &         0.0 &  NIST  \\
   Ti I &         1606.922$^*$&      3000 &      3300 &      $-$0.1 &     $-$0.1 &       $-$0.2 &       $-$0.2 &  NIST  \\
   Ti I &         1635.654$^*$&      3000 &      3300 &      $-$0.3 &        0.1 &       $-$0.3 &         0.2 &  NIST  \\
   Ti I &         1829.502 &      3300 &      3500 &      $-$0.0 &        0.2 &       $-$0.0 &         0.2 &  NIST  \\
   Ti I &         1993.948 &      3300 &      3700 &      $-$0.2 &        0.1 &       $-$0.4 &         0.1 &  NIST  \\
   Ti I &         2201.051 &      3100 &      3700 &      $-$0.2 &     $-$0.0 &       $-$0.4 &       $-$0.1 &  BT-Settl  \\
   Ti I &         2231.672 &      3300 &      3700 &      $-$0.2 &        0.1 &       $-$0.4 &         0.0 &  NIST  \\
   Ti I &         2324.308 &      3200 &      3700 &      $-$0.2 &        0.2 &       $-$0.4 &         0.2 &  NIST  \\
   Ti I &         2328.639 &      3200 &      3700 &      $-$0.2 &        0.2 &       $-$0.4 &         0.2 &  NIST  \\
   Ti I &         2428.846 &      3200 &      3700 &      $-$0.2 &        0.2 &       $-$0.4 &         0.2 &  NIST  \\
   Al I &         1125.628 &      3100 &      3700 &      $-$0.3 &        0.2 &       $-$0.4 &         0.2 &  BT-Settl  \\
   Al I &         1672.353$^*$&      3000 &      3700 &      $-$0.3 &        0.2 &       $-$0.4 &         0.2 &  BT-Settl  \\
   Al I &         1675.513 &      3100 &      3700 &      $-$0.3 &        0.2 &       $-$0.4 &         0.2 &  BT-Settl  \\
   Al I &         2109.875 &      3300 &      3600 &      $-$0.1 &        0.2 &       $-$0.1 &         0.2 &  BT-Settl  \\
   Na I &         1083.787 &      3100 &      3700 &      $-$0.3 &        0.2 &       $-$0.4 &         0.2 &  NIST  \\
   Na I &         1232.334 &      3100 &      3600 &      $-$0.2 &        0.2 &       $-$0.4 &         0.2 &  NIST  \\
   Na I &         1637.832 &      3300 &      3500 &         0.0 &        0.1 &       $-$0.0 &         0.2 &  NIST  \\
    C I &         1133.147 &      3100 &      3300 &      $-$0.1 &     $-$0.0 &       $-$0.2 &      $-$0.1 &  NIST  \\
    C I &         1142.552 &      3300 &      3700 &      $-$0.2 &        0.1 &       $-$0.4 &         0.1 &  NIST  \\
    C I &         1440.391 &      3100 &      3200 &      $-$0.2 &     $-$0.0 &       $-$0.4 &      $-$0.1 &  NIST  \\
    C I &         1454.645 &      3100 &      3300 &      $-$0.0 &        0.0 &       $-$0.1 &         0.0 &  NIST  \\
    C I &         1661.194$^*$&      3200 &      3500 &      $-$0.2 &        0.0 &       $-$0.4 &      $-$0.1 &  NIST  \\
    C I &         1779.812 &      3100 &      3400 &      $-$0.0 &        0.0 &       $-$0.1 &         0.0 &  NIST  \\
    C I &         1803.105 &      3200 &      3500 &      $-$0.0 &        0.0 &       $-$0.1 &         0.0 &  NIST  \\
    C I &         2293.510 &      3000 &      3700 &      $-$0.3 &        0.2 &       $-$0.4 &         0.2 &  NIST  \\
    C I &         2318.485 &      3100 &      3200 &      $-$0.0 &     $-$0.0 &       $-$0.1 &      $-$0.1 &  NIST  \\
    C I &         2354.499 &      3100 &      3600 &      $-$0.3 &        0.1 &       $-$0.4 &         0.1 &  NIST  \\
    K I &         1102.286 &      3200 &      3600 &      $-$0.5 &        0.4 &       $-$0.3 &         0.2 &  NIST  \\
    K I &         1516.721 &      3200 &      3600 &      $-$0.3 &        0.4 &       $-$0.3 &         0.2 &  NIST  \\
    K I &         1517.252 &      3200 &      3600 &      $-$0.3 &        0.4 &       $-$0.3 &         0.2 &  NIST  \\
     OH &         1393.282 &      3200 &      3500 &      $-$0.3 &        0.2 &       $-$0.3 &         0.2 &  BT-Settl  \\
     OH &         1403.609 &      3200 &      3400 &      $-$0.1 &        0.2 &       $-$0.2 &         0.2 &  BT-Settl  \\
     OH &         1408.647 &      3300 &      3400 &      $-$0.1 &        0.2 &       $-$0.2 &         0.2 &  BT-Settl  \\
     OH &         1410.692 &      3300 &      3700 &      $-$0.2 &        0.1 &       $-$0.4 &         0.1 &  BT-Settl  \\
     OH &         1416.298 &      3200 &      3600 &      $-$0.2 &        0.1 &       $-$0.4 &         0.1 &  BT-Settl  \\
     OH &         1418.596 &      3200 &      3400 &      $-$0.2 &        0.1 &       $-$0.4 &         0.0 &  BT-Settl  \\
     OH &         1422.714 &      3300 &      3700 &      $-$0.2 &        0.1 &       $-$0.4 &         0.2 &  BT-Settl  \\
     OH &         1434.455 &      3300 &      3700 &      $-$0.2 &        0.2 &       $-$0.4 &         0.2 &  BT-Settl  \\
     OH &         1435.633 &      3100 &      3600 &      $-$0.3 &        0.2 &       $-$0.4 &         0.2 &  BT-Settl  \\
     OH &         1442.449 &      3200 &      3700 &      $-$0.3 &        0.0 &       $-$0.4 &         0.1 &  BT-Settl  \\
     OH &         1446.921 &      3200 &      3600 &      $-$0.2 &        0.2 &       $-$0.4 &         0.2 &  BT-Settl  \\
     OH &         1456.397 &      3300 &      3700 &      $-$0.2 &        0.0 &       $-$0.4 &      $-$0.0 &  BT-Settl  \\
     OH &         1460.835 &      3200 &      3700 &      $-$0.2 &        0.2 &       $-$0.4 &         0.2 &  BT-Settl  \\
     OH &         1462.687 &      3200 &      3600 &      $-$0.1 &        0.1 &       $-$0.4 &         0.0 &  BT-Settl  \\
     OH &         1469.887 &      3100 &      3700 &      $-$0.3 &        0.2 &       $-$0.4 &         0.2 &  BT-Settl  \\
     OH &         1479.923 &      3300 &      3500 &      $-$0.0 &        0.2 &       $-$0.0 &         0.2 &  BT-Settl  \\
     OH &         1480.042 &      3300 &      3700 &      $-$0.2 &        0.2 &       $-$0.4 &         0.2 &  BT-Settl  \\
     OH &         1490.978 &      3200 &      3600 &      $-$0.1 &        0.2 &       $-$0.3 &         0.2 &  BT-Settl  \\
     OH &         1500.722 &      3100 &      3500 &      $-$0.3 &        0.1 &       $-$0.4 &         0.2 &  BT-Settl  \\
     OH &         1502.696 &      3300 &      3700 &      $-$0.2 &        0.1 &       $-$0.4 &         0.1 &  BT-Settl  \\
     OH &         1505.320 &      3200 &      3300 &      $-$0.2 &        0.1 &       $-$0.4 &         0.1 &  BT-Settl  \\
     OH &         1505.566 &      3200 &      3700 &      $-$0.2 &        0.1 &       $-$0.4 &         0.1 &  BT-Settl  \\
     OH &         1526.878 &      3100 &      3700 &      $-$0.2 &        0.2 &       $-$0.4 &         0.2 &  BT-Settl  \\
     OH &         1528.524 &      3200 &      3500 &      $-$0.2 &        0.0 &       $-$0.4 &      $-$0.1 &  BT-Settl  \\
     OH &         1533.212 &      3300 &      3600 &         0.0 &        0.2 &       $-$0.0 &         0.2 &  BT-Settl  \\
     OH &         1537.919 &      3300 &      3700 &      $-$0.2 &        0.2 &       $-$0.4 &         0.2 &  BT-Settl  \\
     OH &         1538.694 &      3300 &      3700 &      $-$0.2 &        0.2 &       $-$0.4 &         0.2 &  BT-Settl  \\
     OH &         1538.925 &      3200 &      3500 &      $-$0.1 &        0.1 &       $-$0.2 &         0.1 &  BT-Settl  \\
     OH &         1539.525 &      3100 &      3600 &      $-$0.2 &        0.1 &       $-$0.3 &         0.1 &  BT-Settl  \\
     OH &         1540.717 &      3100 &      3700 &      $-$0.2 &        0.2 &       $-$0.4 &         0.2 &  BT-Settl  \\
     OH &         1541.339 &      3200 &      3300 &      $-$0.3 &        0.1 &       $-$0.4 &         0.1 &  BT-Settl  \\
     OH &         1556.228 &      3500 &      3600 &      $-$0.2 &        0.0 &       $-$0.2 &      $-$0.0 &  BT-Settl  \\
     OH &         1557.007 &      3300 &      3700 &      $-$0.2 &        0.1 &       $-$0.4 &         0.1 &  BT-Settl  \\
     OH &         1563.095 &      3000 &      3500 &      $-$0.3 &        0.0 &       $-$0.4 &         0.0 &  BT-Settl  \\
     OH &         1563.168 &      3000 &      3300 &      $-$0.3 &     $-$0.0 &       $-$0.3 &      $-$0.1 &  BT-Settl  \\
     OH &         1565.778 &      3100 &      3600 &      $-$0.1 &        0.0 &       $-$0.3 &         0.0 &  BT-Settl  \\
     OH &         1583.314 &      3300 &      3700 &      $-$0.2 &        0.2 &       $-$0.4 &         0.2 &  BT-Settl  \\
     OH &         1589.219 &      3300 &      3600 &      $-$0.1 &        0.1 &       $-$0.3 &         0.0 &  BT-Settl  \\
     OH &         1591.707 &      3100 &      3500 &      $-$0.3 &        0.2 &       $-$0.4 &         0.2 &  BT-Settl  \\
     OH &         1604.126 &      3200 &      3700 &      $-$0.3 &        0.2 &       $-$0.4 &         0.2 &  BT-Settl  \\
     OH &         1606.943$^*$&      3200 &      3500 &      $-$0.2 &        0.0 &       $-$0.2 &      $-$0.1 &  BT-Settl  \\
     OH &         1607.968 &      3100 &      3600 &      $-$0.2 &        0.1 &       $-$0.4 &         0.1 &  BT-Settl  \\
     OH &         1619.454 &      3200 &      3500 &      $-$0.2 &        0.0 &       $-$0.4 &      $-$0.1 &  BT-Settl  \\
     OH &         1621.162 &      3200 &      3500 &      $-$0.3 &        0.2 &       $-$0.4 &         0.2 &  BT-Settl  \\
     OH &         1625.944 &      3300 &      3700 &      $-$0.2 &        0.2 &       $-$0.4 &         0.2 &  BT-Settl  \\
     OH &         1635.670$^*$&      3000 &      3300 &      $-$0.3 &        0.0 &       $-$0.3 &         0.0 &  BT-Settl  \\
     OH &         1644.796 &      3300 &      3700 &      $-$0.2 &        0.2 &       $-$0.4 &         0.2 &  BT-Settl  \\
     OH &         1647.734 &      3000 &      3700 &      $-$0.3 &        0.1 &       $-$0.4 &         0.1 &  BT-Settl  \\
     OH &         1665.450 &      3000 &      3300 &      $-$0.3 &     $-$0.0 &       $-$0.3 &       $-$0.1 &  BT-Settl  \\
     OH &         1669.226 &      3300 &      3600 &      $-$0.1 &        0.1 &       $-$0.3 &         0.1 &  BT-Settl  \\
     OH &         1672.342$^*$&      3100 &      3700 &      $-$0.3 &        0.1 &       $-$0.4 &         0.1 &  BT-Settl  \\
     OH &         1675.385$^*$&      3100 &      3700 &      $-$0.3 &        0.2 &       $-$0.4 &         0.2 &  BT-Settl  \\
     OH &         1675.630 &      3200 &      3700 &      $-$0.2 &        0.2 &       $-$0.4 &         0.2 &  BT-Settl  \\
     OH &         1688.913$^*$&      3100 &      3700 &      $-$0.3 &        0.2 &       $-$0.4 &         0.2 &  BT-Settl  \\
     OH &         1690.339$^*$&      3100 &      3600 &      $-$0.3 &        0.1 &       $-$0.4 &         0.0 &  BT-Settl  \\
     OH &         1705.688$^*$&      3000 &      3300 &      $-$0.2 &        0.0 &       $-$0.4 &      $-$0.0 &  BT-Settl  \\
     OH &         1711.495 &      3300 &      3700 &      $-$0.2 &        0.1 &       $-$0.4 &         0.1 &  BT-Settl  \\
     OH &         1717.982 &      3000 &      3600 &      $-$0.3 &        0.1 &       $-$0.3 &         0.1 &  BT-Settl  \\
     OH &         1721.010&      3000 &      3300 &      $-$0.3 &     $-$0.0 &       $-$0.4 &      $-$0.1 &  BT-Settl  \\
     OH &         1733.871$^*$&      3300 &      3700 &      $-$0.2 &        0.2 &       $-$0.4 &         0.2 &  BT-Settl  \\
     OH &         1781.951 &      3100 &      3600 &      $-$0.3 &        0.0 &       $-$0.4 &         0.0 &  BT-Settl  \\
     OH &         1782.337 &      3200 &      3600 &      $-$0.1 &        0.1 &       $-$0.4 &         0.0 &  BT-Settl  \\
     OH &         1801.969 &      3300 &      3700 &      $-$0.2 &        0.2 &       $-$0.4 &         0.2 &  BT-Settl  \\
\end{longtable}
Note — Wavelengths marked with an asterisk (*) indicate the presence of nearby spectral features based on our line lists. Nearby lines suggest proximity to our target lines, but their influence on the target lines' strength is unclear. Our previous empirical tests showed no significant differences in abundance between the full list and those without contaminated ones (\citealt{Jahandar2023}).

\clearpage

\section{Distribution histogram of different elements} \label{sec:p2_appendix2}

\begin{figure}[ht]
    \centering
    \includegraphics[width=1\linewidth]{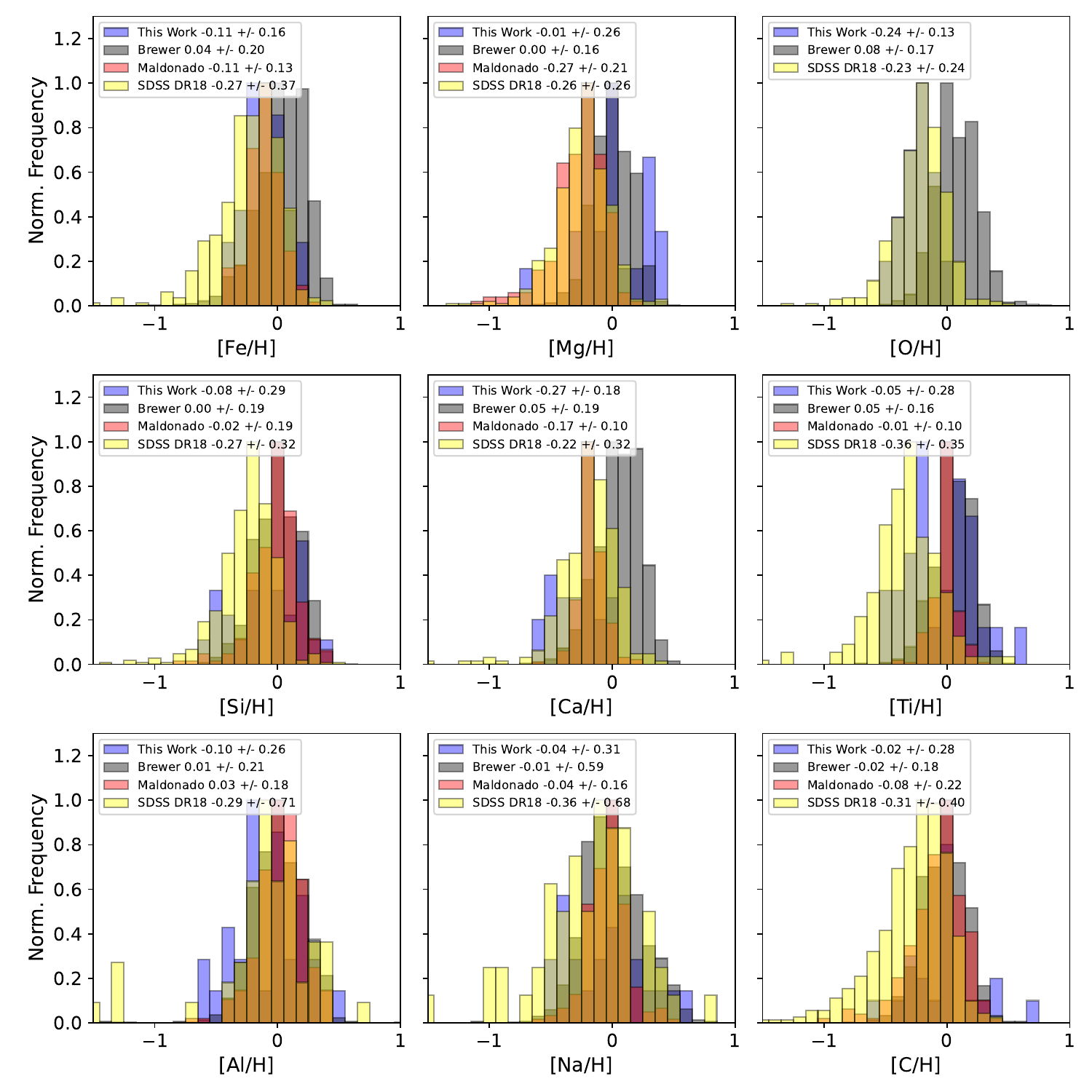}
    \caption{Comparison between the chemistry of M dwarfs from this work, the SDSS database (DR18), \citealt{maldonado2020hades} and FGK stars from \cite{brewer2016spectral}. 
    \label{fig:hist_all_el} }
\end{figure}

Figure~\ref{fig:hist_all_el} provide histograms comparing the distribution of various elemental abundances — [Fe/H], [Mg/H], [O/H], [Si/H], [Ca/H], [Ti/H], [Al/H], [Na/H], and [C/H] — for our M dwarf sample, alongside those of M dwarfs from \cite{maldonado2020hades} and SDSS DR18, as well as FGK stars from \cite{brewer2016spectral}. Note that we excluded the [K/H] comparison, as it was only available in our study and SDSS DR18 (see Figure \ref{fig:master_chemistry}).

\clearpage

\end{document}